\documentclass{jfm}
\usepackage{amssymb}
\usepackage{bm}
\usepackage{natbib}

\usepackage{graphicx}
\usepackage{amsmath}
\usepackage{verbatim}
\usepackage{color}
\usepackage{subfigure}

\ifCUPmtlplainloaded \else
  \checkfont{eurm10}
  \iffontfound
    \IfFileExists{upmath.sty}
      {\typeout{^^JFound AMS Euler Roman fonts on the system,
                   using the 'upmath' package.^^J}%
       \usepackage{upmath}}
      {\typeout{^^JFound AMS Euler Roman fonts on the system, but you
                   dont seem to have the}%
       \typeout{'upmath' package installed. JFM.cls can take advantage
                 of these fonts,^^Jif you use 'upmath' package.^^J}%
      }
  \else
  \fi
\fi

\ifCUPmtlplainloaded \else
  \checkfont{msam10}
  \iffontfound
    \IfFileExists{amssymb.sty}
      {\typeout{^^JFound AMS Symbol fonts on the system, using the
                'amssymb' package.^^J}%
       \usepackage{amssymb}%
         \let\leq=\leqslant
         \let\geq=\geqslant
      }{}
  \fi
\fi

\usepackage{hyperref}
\hypersetup{
	hyperindex,
	breaklinks,
	colorlinks=true,
	linkcolor=blue,
	citecolor=blue,
	bookmarks=true,
	bookmarksopen=true,
	bookmarksopenlevel=2,
	pdfstartpage={1},
	pdfstartview={FitH},
	pdfview={FitH 0},
	pdfauthor={Nikolaos A. Bakas and Petros J. Ioannou},
	pdftitle={A theory for the emergence of coherent structures in beta-plane turbulence},
 }
\usepackage[all]{hypcap}	

\ifCUPmtlplainloaded \else
  \IfFileExists{amsbsy.sty}
    {\typeout{^^JFound the 'amsbsy' package on the system, using it.^^J}%
     \usepackage{amsbsy}}
    {}
\fi

\renewcommand{\theequation}{\thesection.\arabic{equation}}

\newcommand{\md}{\mathrm{d}}
\newcommand{\me}{\mathrm{e}}

\newcommand{\qq}{\begin{eqnarray}}
\newcommand{\qqq}{\end{eqnarray}}

\begin{document}
	
\title{Fluctuations of large-scale jets in the stochastic 2D Euler equation}

\author[Cesare Nardini and Tom\'as Tangarife]%
{Cesare Nardini$^{1,2}$ %
  \thanks{Email address for correspondence: cesare.nardini@gmail.com},\ns
and Tom\'as Tangarife$^3$}

\affiliation{$^1$DAMTP, Centre for Mathematical Sciences, University of Cambridge, Wilberforce Road, Cambridge CB3 0WA, UK\\[\affilskip]
$^2$SUPA, School of Physics and Astronomy, University of Edinburgh, Peter Guthrie Tait Road, Edinburgh EH9 3FD, UK\\[\affilskip]
$^3$Laboratoire de Physique de l'Ecole Normale Sup\'erieure 
de Lyon, Universit\'e de Lyon and CNRS, 46, All\'ee d'Italie, 
F-69007 Lyon, France}

\maketitle

\begin{abstract}
Two-dimensional turbulence in a rectangular domain self-organises into large-scale unidirectional jets. While several results are present to characterize the mean jets velocity profile, much less is known about the fluctuations. We study jets dynamics in the stochastically forced two-dimensional Euler equations. In the limit where the average jets velocity profile evolves slowly with respect to turbulent fluctuations, we employ a multi-scale (kinetic theory) approach, which relates jet dynamics to the statistics of Reynolds stresses. We study analytically the Gaussian fluctuations of Reynolds stresses and predict the spatial structure of the jets velocity covariance. 
Our results agree qualitatively well with direct numerical simulations, clearly showing that the jets velocity profile are enhanced away from the stationary points of the average velocity profile.
A numerical test of our predictions at quantitative level seems out of reach at the present day. 
\end{abstract}


\section{Introduction}

Turbulence in planetary atmospheres, oceans \citep{vallisatmospheric2006}, rotating flows \citep{morize2005decaying} and two-dimensional turbulence \citep{sommeria1986,Yin_Montgomery_Clercx_2003PhFluids} leads very often to self-organisation
in large scale coherent structures. The explanation of their emergence and their
characterisation is a major theoretical issue in atmospheric and oceanic dynamics as well a very challenging question from a fundamental point of view. One of the commonly observed large scale structures are jets, i.e. flows that are in average orizontal and unidirectional; the reader can consult the special issue of
Journal of Atmospherical Science \citep{JOAjetsspecial} for a recent account on the jet dynamics in planetary atmospheres and oceans. A similar self-organization into
jets has also been observed in two-dimensional turbulence \citep{Bouchet_Simonnet_2008,Yin_Montgomery_Clercx_2003PhFluids}.

Much effort has been devoted in literature to characterise the average structure of jets, such as their mean velocity profile as a function of the latitude. Little is instead known on the fluctuations, small and large, they undergo. It is however of obvious crucial importance to understand how far the instantaneous jet velocity profiles typically are from their average state. The present paper deals with such a question in a very simple, yet relevant, setup.



The aim of this work is to consider an approach based on statistical
mechanics to study the fluctuations of the jet velocity profile.  
We will focus on a non-equilibrium statistical mechanics approach: forces and dissipation induce a flow of energy from small to large scales where it is dissipated.
 The theoretical framework we employ has been originally developed in \citep{Bouchet_Nardini_Tangarife_2013_Kinetic,nardini2014fluiddyn} and shares strong similarities with: i) kinetic theories for plasma and gravitational systems \citep{Nicholson_1991,balescu1975equilibrium,Nardini_Gupta_Ruffo_Dauxois_Bouchet_2012_kinetic,NardiniGuptaBouchet-2012-JSMTE,heyvaerts2010balescu}; ii) theories based on quasi-linear approximation such as second-order cumulant expansion (CE2: \citep{Marston-2010-Chaos,Marston-APS-2011Phy,tobias2013direct,Marston_Conover_Schneider_JAS2008,Marston-APS-2011Phy,GormanSchneider-QL-GCM,Srinivasan-Young-2011-JAS,ait2015cumulant}) and Stochastic Structural Stability Theory (SSST: \citep{Farrel_Ioannou,Farrell_Ioannou_JAS_2007,BakasIoannou2013SSST,ParkerKrommes2013SSST}); iii) averaging and homogenization theory for stochastic systems \citep{Gardiner_1994_Book_Stochastic,pavliotis2008multiscale}. In few words, our theory is a multi-scale approach: an effective dynamics of the jet velocity profile is derived integrating out turbulent non-zonal fluctuations. 
 
 Even if similar to CE2 at first sight, our multi-scale approach differs from them and is expected to capture not only the evolution of the average jet velocity profile but also its Gaussian fluctuations. In this paper we show for the first time that this is indeed the case: we analytically derive predictions for the fluctuations of the jet velocity profile close to its steady state and test them against direct numerical simulations. We believe this to be an important result, as it is well known \citep{kraichnan1980realizability,marston2014directbook} that improvement of the cumulant expansion beyond second order suffers of realizability problems: these approaches are not self-consistent without further ad-hoc approximations (see however \citep{marston2014directbook} for a recent careful discussion of this point and implementations of the cumulant expansion beyond second order). 

In order to progress as far as possible with analytical tools, we consider in this paper the stochastically forced $2$-d Euler equation.
We analyse the covariance and variance of the jet velocity profile, defined as
\qq\label{eq:covariance-variance-intro}
\mathbb{C}(y_1,y_2,\infty) = \lim_{t\to\infty}\mathbb{E}[U(y_1,t) \, U(y_2,t)] -U_d(y_1,t) \,U_d(y_2,t) \qquad \qquad 
\mathbb{V}(y,\infty) = \lim_{t\to\infty}\mathbb{C}(y,y,t)\,,
\qqq
where $U(y,t)$ is the jet velocity profile (i.e., the zonal component of the zonally averaged velocity profile) and $U_d$ is the one averaged over noise realisations. We predict their spatial structure, showing that $\mathbb{C}(y_1,y_2,\infty)$ is highly enhanced at those points where $U_d(y_1)=U_d(y_2)$, except if $U_d'(y_1)=0$. Thus, the jet velocity profiles fluctuates much more away from the stationary points of $U_d(y)$ than close to them.\\
More precisely,
denoting by $\alpha$ the ratio between i) the time-scale for the advection of small scales vortices by the large scale jet and ii) the time-scale for the evolution of $U_d$, we consider the limit where (hyper)-viscosity is negligible and $\alpha $ is a small parameter. If we were dealing with a system with a finite number of degrees of freedom, one would expect that both the variance and the covariance of the zonal jet velocity profile were small, of order $\mathcal{O}(\alpha)$. However, in the context of the $2$D stochastic Euler equations, ultraviolet divergences renormalise such result. 
We will indeed show that, 
in proper non-dimensional units that will be introduced in section \ref{sec:slow_jet_dynamics}, our theory predicts that $\mathbb{C}(y_1,y_2,\infty) \sim \mathcal{O}(\alpha)$ except when $U_d(y_1)=U_d(y_2)$ and $U_d'(y_1)\neq 0$, where we have $\mathbb{C}(y_1,y_2,\infty) \sim \mathcal{O}(1)$. Moreover, $\mathbb{V}(y,\infty) \sim \mathcal{O}(1)$ unless $U_d'(y)=0$, in which case $\mathbb{V}(y,\infty) \sim \mathcal{O}(\alpha)$. Mathematically, we can express such result saying that $\mathbb{C}(y_1,y_2,\infty)$ (but not $\mathbb{V}(y,\infty)$ !) converges in distributional sense.

Employing direct numerical simulations, we find a clear footprint of out predictions in the behaviour of the fully non-linear stochastic 2D Euler equations: $\mathbb{C}(y_1,y_2,\infty)$ and $\mathbb{V}(y,\infty)$ present a spatial structure very similar to the one described above. The reader can already consult figures \ref{fig:UU_vs_CV_lx07_beta0_N256} and \ref{fig:meanU_vs_varU_lx07_beta0_N256} in this respect. Unfortunately, a quantitative comparison of our theoretical results with direct numerical simulations seems out of reach at the present day. In particular, we are unable to verify the correctness of the above scaling with $\alpha$: the problem is  computationally  hard, as one has to work in the limit of negligible (hyper)-viscosity and very slow evolution.


The paper is organised as follows. In section \ref{sec:slow_jet_dynamics} we briefly recap the theoretical framework originally developed in \citep{Bouchet_Nardini_Tangarife_2013_Kinetic}: the effective evolution equation for the jet velocity profile is given in \eqref{eq:kinetic_equation} and is a stochastic partial differential equation. In section \ref{sec:Fluctuations_Reynolds} we compute the statistical properties of the noise for an arbitrary average jet velocity profile. This section contains the central theoretical result of this paper which is  summarised in eq. \eqref{eq:SC-Xi-Xi_kl} and (\ref{eq:SC-Xi-kl-convergence}): it describes the spatial structure of the covariance of the noise appearing in the effective evolution of the jet velocity profile. 
In section \ref{sec:Consequences_mean_flow}, we discuss the implication for the variance and the covariance of the jet velocity profile and compare our predictions to direct numerical simulations of stochastically forced 2D Euler equation. We conclude in section \ref{sec:Conclusion} with a summary and the perspective of our work. Several appendices contain those results that are too technical to fit in the body of the paper.

\section{Two-dimensional stochastic Euler dynamics in a rectangular domain\label{sec:slow_jet_dynamics}}
We consider the dynamics of a two-dimensional (2D) flow subjected to random forces, described by the equation
\begin{equation}
\frac{\partial \omega}{\partial t} + \mathbf{v}\cdot\nabla \omega = -\lambda \omega -  \nu_n \left(-\Delta\right)^n \omega + \sqrt{\sigma}\eta\,,
\label{eq:2D-NS-dimensional}
\end{equation}
for the vorticity field $\omega(\mathbf{r},t)$, where $\mathbf{r}=(x,y)$ is the two-dimensional position vector. Following the language commonly used in geophysical fluid dynamics, we refer to $x$ as the zonal and to $y$ as the meridional directions.
 The velocity field $\mathbf{v}(\mathbf{r},t)=(u,v)(\mathbf{r},t)$ is related to the vorticity field through the relation $\omega = \left(\nabla\times\mathbf{v}\right)\cdot\mathbf{e}_z$, where $\mathbf{e}_z$ denotes the unit vector normal to the surface of the flow. 
 Energy is dissipated at large scales by  Rayleigh (or Ekman) friction with coefficient
 $\lambda$ and $\nu_{n}$
is the hyper-viscosity coefficient (viscosity for $n=1$). When $n=1$ equation \eqref{eq:2D-NS-dimensional} is the 2D Navier-Stokes equation; however as we will be particularly interested in the limit in which viscosity is negligible and only serve to stabilise the numerical integration of \eqref{eq:2D-NS-dimensional},
we just refer to  equation \eqref{eq:2D-NS-dimensional} as the 2D Euler equation.

We consider the case of a biperiodic domain $\mathcal{D}=[0,2\pi L/r)\times[0,2\pi L r)$ with aspect
ratio $r^2$, i.e. the velocity and vorticity fields satisfy $f\left(x+2\pi L/r,y\right)=f\left(x,y\right)$
and $f\left(x,y+2\pi Lr\right)=f\left(x,y\right)$. The forcing term $\eta$ is assumed to be a white in time Gaussian noise
with autocorrelation function 
\qq
\mathbb{E}\left[\eta(\mathbf{r}_{1},t_{1})\eta(\mathbf{r}_{2},t_{2})\right]=C(\mathbf{r}_{1}-\mathbf{r}_{2})\delta(t_{1}-t_{2})
\qqq
where $\mathbb{E}$ is the expectation over realizations of the noise $\eta$; the covariance $C$ is required to be an even positive definite function, periodic with respect
to $x$ and $y$. As discussed below, $\sigma$ is the average energy
input rate. We assume in this paper that the noise autocorrelation
function $C$ is translationally invariant in both direction. The
fact that $C$ is invariant in the $x$-direction is important for some of the computations. However the generalization to non-meridionally invariant
forcing would be straightforward. Moreover, we do not force directly the jet, by imposing that 
\qq
C_z\equiv
\frac{1}{2\pi/r}\int_{0}^{2\pi/ r} C({\bf r})\,\mbox{d}x. =0\,.
\qqq
This setup is sounding for many applications, as forcing typically acts at scales much smaller than the ones where coherent structures develop.

In the turbulent regime that will be defined later on, the stochastic 2D Euler dynamics \eqref{eq:2D-NS-dimensional} self-organises into long-living coherent structures at the large scales of the flow \citep{BouchetVenaille-PhysicsReport}. In a square domain (with aspect ratio $r=1$), this large-scale structure is a dipole of vortices. In a rectangular domain with $r>1$ ($r<1$), the large-scale structure is a parallel flow in the $x$ ($y$) direction. The situation is similar to zonal jet formation in models of geophysical turbulent flows \citep{KraichnanMontgomery1980,vallisatmospheric2006}. 

\subsection{Energy balance and non--dimensional equations\label{sub:energy-balance}}

Equation (\ref{eq:2D-NS-dimensional}) with $\lambda=\sigma=\nu_{n}=0$
describes a perfect 2D flow. The equations are then Hamiltonian
and they conserve the kinetic energy
\begin{equation}
\mathcal{E}[\omega]=\frac{1}{2}\int_{\mathcal{D}}\mbox{d}\mathbf{r}\:\mathbf{v}^{2}=-\frac{1}{2}\int_{\mathcal{D}}\mbox{d}\mathbf{r}\:\omega\Delta^{-1}\omega
\end{equation}
where $\Delta^{-1}$ denotes the inverse Laplacian operator; and the Casimir functionals
\begin{equation}
\mathcal{C}_{s}[\omega]=\int_{\mathcal{D}}\mbox{d}\mathbf{r}\: s(\omega),
\end{equation}
for any sufficiently regular function $s$.

Because the force $\eta$ in (\ref{eq:2D-NS-dimensional}) is a white in time Gaussian process, we can compute
a-priori the average, with respect to noise realizations, of the input
rate for quadratic invariants. We impose, without loss of generality (indeed, multiplying $C$ by an arbitrary positive constant amounts at renormalizing
$\sigma$) that
that 
\[
-2\pi^{2}L^{2}\left(\Delta^{-1}C\right)(\mathbf{0})=1\,.
\]
With the above choice, the average energy input rate
is $\sigma$ and the average energy input rate by unit of area is
$\epsilon=\sigma/4\pi^{2}L^{2}$.
We then  consider the energy balance for equation (\ref{eq:2D-NS-dimensional}),
with $E=\mathbb{E}\left[\mathcal{E}[\omega]\right]$:
\begin{equation}
\frac{dE}{dt}=-2\lambda E-\nu_{n}H_{n}+\sigma,
\label{eq:Energy-Balance}
\end{equation}
where $H_{n}=-\mathbb{E}\left[\int_{\mathcal{D}}\psi\left(-\Delta\right)^{n}\omega\right]$.
For most of the turbulent flows we are interested in, the ratio $2\lambda E/\nu_{n}H_{n}$
will be extremely large  and (hyper)-viscosity is negligible for energy dissipation.
Then, in a statistically stationary regime, the approximate average
energy is $E\simeq\sigma/2\lambda$,
expressing the balance between stochastic forces and linear friction
in \eqref{eq:2D-NS-dimensional}. This average total energy estimate yields the typical jet velocity
$U\sim\sqrt{\sigma/2\lambda}$, so that an estimate of the 
time scale for the advection 
of small-scale turbulent vortices by the large scale jet
is $\tau\sim L /U$. 

We thus perform a transformation to
non-dimensional variables such that in the new units the domain is
$\mathcal{D}=[0,2\pi /r)\times[0,2\pi r)$ and the approximate average
energy is 1. This is done introducing a non-dimensional time $t'=t/\tau$
and a non-dimensional spatial variable $\mathbf{r}'=\mathbf{r}/L$
with $\tau=L^{2}\sqrt{2\lambda/\sigma}$. The non-dimensional physical
variables are then $\omega'=\tau \omega$, $\mathbf{v}'=\tau\mathbf{v}/L$,
and the non-dimensional parameters are defined by
\begin{equation}
\alpha=\lambda\tau=L^{2}\sqrt{\frac{2\lambda^{3}}{\sigma}}=\frac{L}{2\pi}\sqrt{\frac{2\lambda^{3}}{\epsilon }},
\label{eq:alpha}
\end{equation}
 $\nu_{n}'=\nu_{n}\tau/L^{2n}=\nu_{n}\sqrt{2\lambda/\sigma}/L^{2n-2}$.
Moreover, a rescaled stochastic Gaussian field $\eta'$ appears, with $\mathbb{E}\left[\eta'(\mathbf{r}'_{1},t'_{1})\eta(\mathbf{r}'_{2},t'_{2})\right]=C'(\mathbf{r}'_{1}-\mathbf{r}'_{2})\delta(t'_{1}-t'_{2})$
with $C'(\mathbf{r}')=L^{4}C(\mathbf{r})$. Performing the adimensionalization
procedure explained above, the 2D Euler equation reads
\begin{equation}
\frac{\partial \omega}{\partial t} + \mathbf{v}\cdot\nabla \omega = -\alpha \omega -  \nu_n \left(-\Delta\right)^n \omega + \sqrt{2\alpha}\eta\,,
\label{eq:2D-NS}
\end{equation}
where, for easiness in the notations, we drop here and in the following
the primes. We note that in non-dimensional units, $\alpha$ represents
an inverse Reynolds number based on the large scale dissipation of
energy and $\nu_{n}$ is an inverse Reynolds number based on the viscosity
or hyper-viscosity term that acts predominantly at small scales. Moreover, phenomenologically, one expects that $\alpha$ is the ratio between the time scale for the evolution of small scales $\tau$ defined above with the time scale of evolution of large-scale coherent structures, given by the dissipative time scale $1/\lambda$.
We address the reader to \citep{nardini2016galpbook} for a longer discussions on the adimensionalization procedure and its generalisation to barotropic flows.

\subsubsection{Numerical simulations and phenomenlogy\label{sub:Numerical_simulations}}
\begin{figure}
\subfigure[\label{fig:hovmoller_a1_lx07_beta0_N256}$\alpha=10^{-3}$]{\includegraphics[height=4cm]{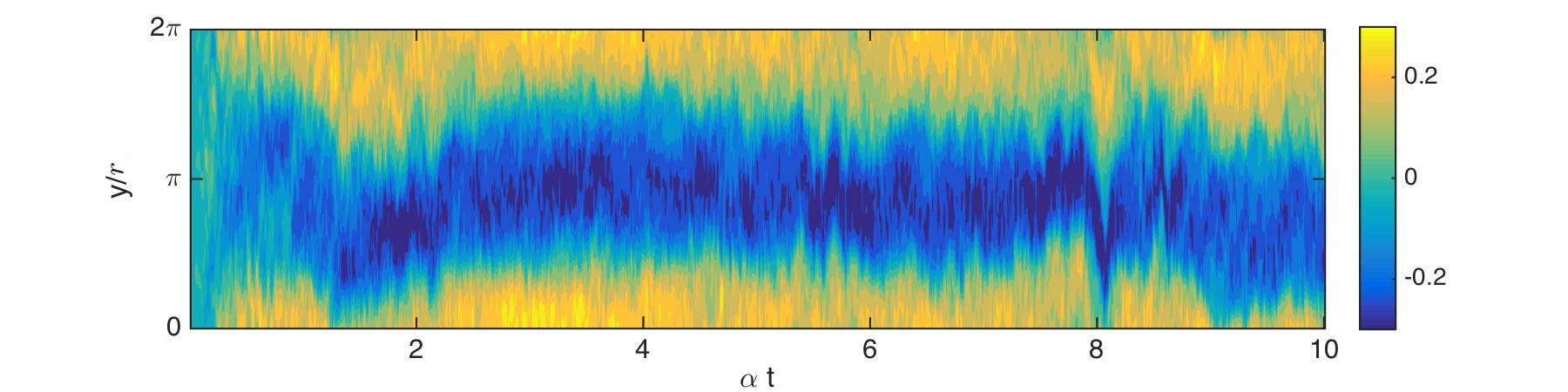}}\\
\subfigure[\label{fig:hovmoller_a05_lx07_beta0_N256}$\alpha=5.10^{-4}$]{\includegraphics[height=4cm]{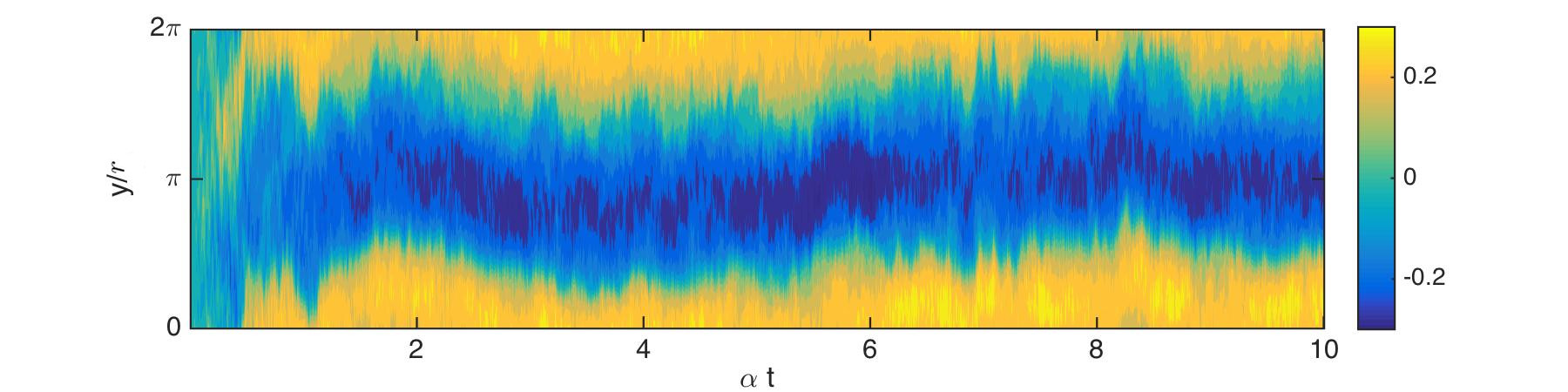}}
\caption{\label{fig:hovmoller_lx07_beta0_N256}
Spatio-temporal diagram of the $x$-component of the velocity averaged over the $x$ direction (Hovmoller diagram of the zonal velocity profile $U(y,t)$), for (a) $\alpha=10^{-3}$ and (b) $\alpha=5.10^{-4}$. After a time $t\sim 2/\alpha$, the flow self-organises into a strong jet in the $x$ direction (zonal jet). This jet undergoes fluctuations in its position, shape and amplitude, see also figure \ref{fig:meanU_Usnapshot_lx07_beta0_N256}. The parameters used in these simulations are detailed in section \ref{sub:Numerical_simulations}.}
\end{figure}

A part of this paper contains the results of direct numerical simulations of the  stochastic 2D Euler equation \eqref{eq:2D-NS} performed employing a pseudo-spectral code. Most of the numerical results shown are obtained at resolution $256\times256$ with hyper-viscosity of order $n=4$ and coefficient $\nu_4=7.10^{-17}$. Examples of such simulations with $\alpha=10^{-3}$ and $\alpha=5.10^{-4}$ are represented in figure \ref{fig:hovmoller_lx07_beta0_N256}. We have checked that our results do not depend on hyper-viscosity by performing few test simulations (data not shown) with resolution $512\times512$ and hyper-viscosity $\nu_4=3,5.10^{-17}$.


All through the paper we use an homogeneous stochastic forcing with spectrum concentrated around wavenumbers $(k,l)$ such that $k^2+l^2 = \left(k_f\pm \delta k\right)^2 $ with $k_f=8$ and $\delta k=1$, except for $k=0$ (i.e. $C_z=0$). The stochastic forcing is generated with Gaussian random numbers which are added to the evolution equation every 10 time steps, using an Euler-Maruyama scheme. 

The numerical results presented in the paper correspond to the stochastic 2D Euler equation \eqref{eq:2D-NS} in a biperiodic domain with aspect ratio $r=1.2$. We have checked that very similar results are obtained, however, using different values of $r>1$.


In this paper, we will study the regime $\nu_n\ll\alpha\ll1$, where the zonal jet velocity profile
\qq\label{eq:U-new}
U(y)\equiv
\frac{1}{2\pi/r}\int_{0}^{2\pi/ r}u(x,y)\,\mbox{d}x.
\qqq
evolves over (non-dimensional) time scales of order $1/\alpha$, while it is forced by Reynolds stresses which evolve over time scales of order 1. 
Direct numerical simulations illustrate this phenomenology: a time-scale separation is indeed present between the time-scale for the jet velocity profile and the one for small turbulent fluctuations. In figure \ref{fig:hovmoller_lx07_beta0_N256} is reported the Hovmoller diagram for the zonal jet velocity profile ($x$-component of the velocity averaged over the $x$ direction) as a function of time, for two different values of $\alpha\ll1$. It is clear that the zonal jet forms at $\alpha t\simeq 1$. 

\begin{figure}
\subfigure[\label{fig:meanU_Usnapshot_a1_lx07_beta0_N256}$\alpha=10^{-3}$]{\includegraphics[height=6cm]{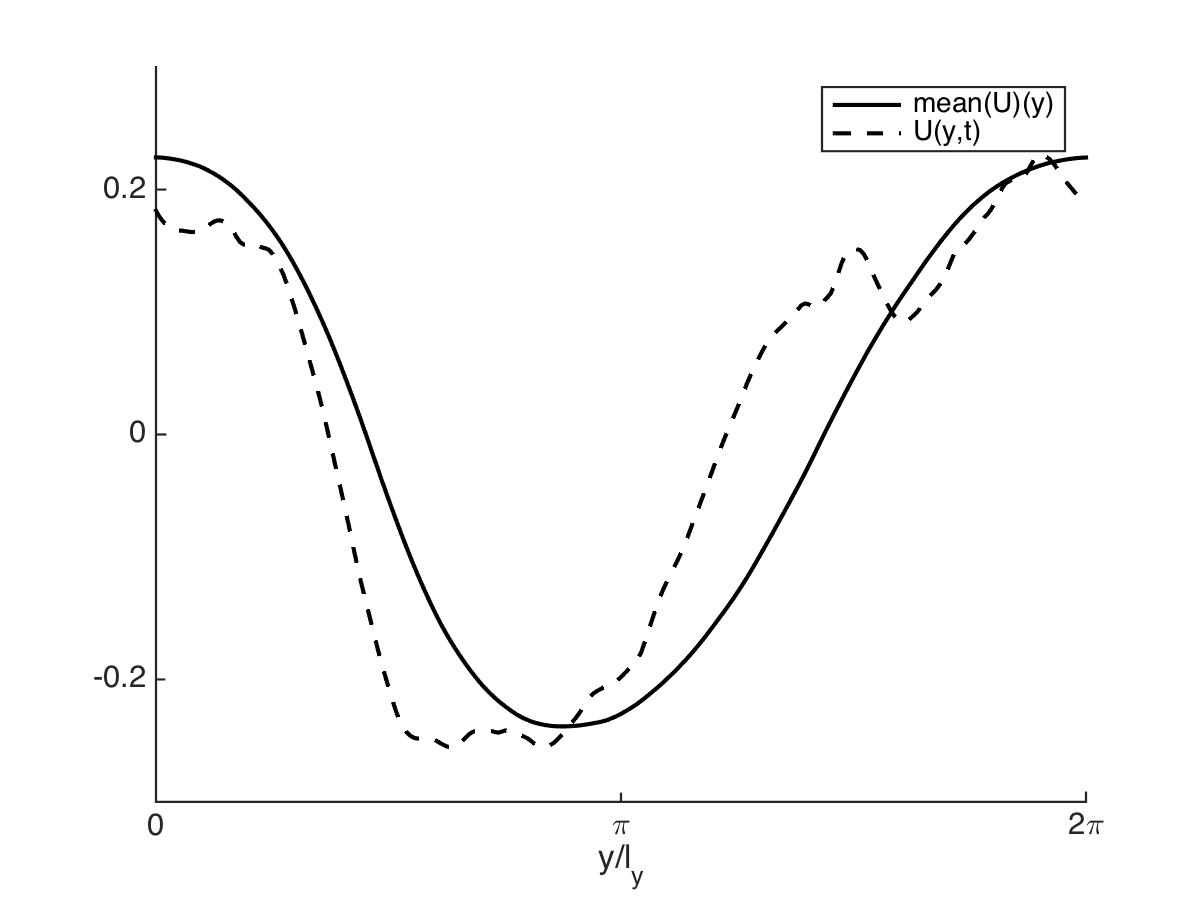}}
\subfigure[\label{fig:meanU_Usnapshot_a05_lx07_beta0_N256}$\alpha=5.10^{-4}$]{\includegraphics[height=6cm]{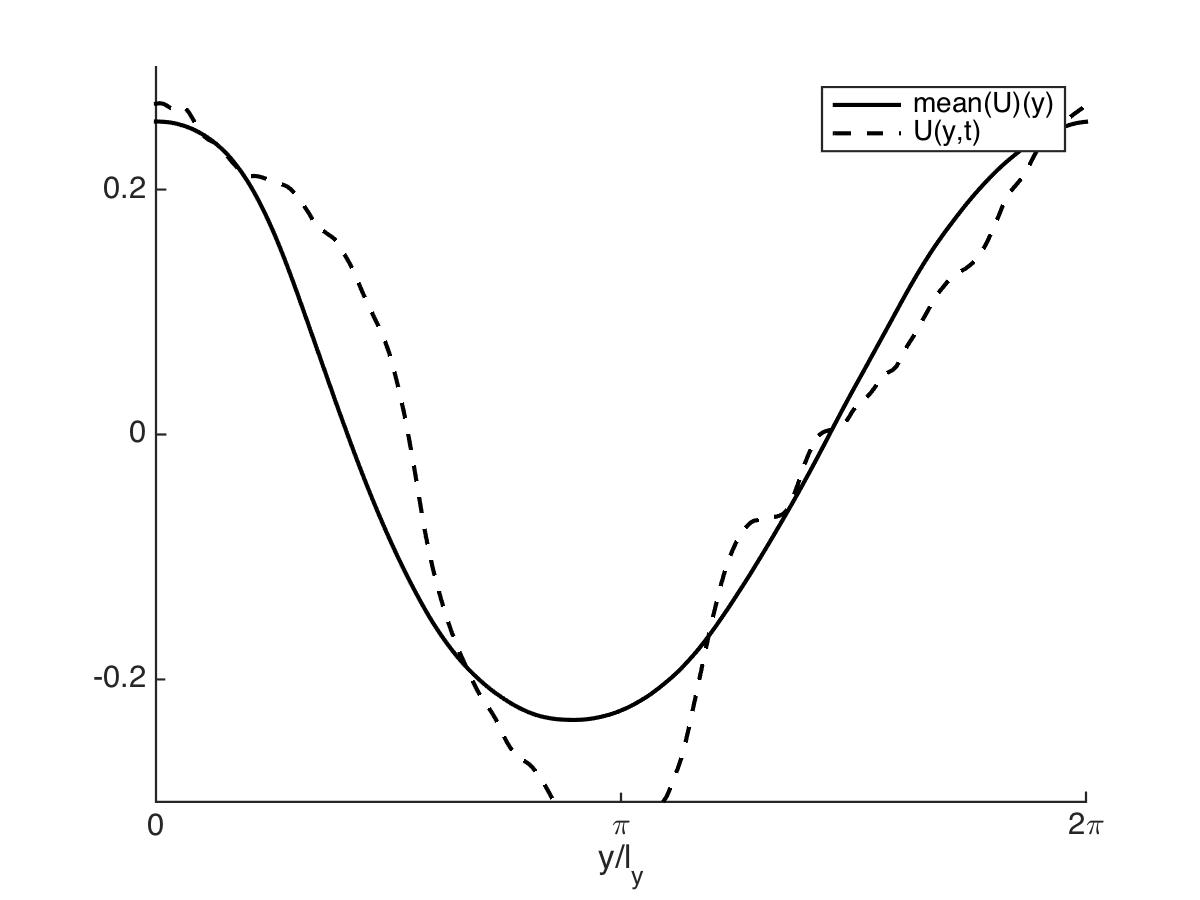}}
\caption{\label{fig:meanU_Usnapshot_lx07_beta0_N256}Mean zonal velocity $\mathbb{E}[U](y)$ (solid curve) and snapshot of the zonal velocity $U(y,t)$ (dashed curve), for (a) $\alpha=10^{-3}$ and (b) $\alpha=5.10^{-4}$. The mean zonal velocity is computed over a period $2/\alpha$ in which the position of the jet is considered to be almost steady, i.e. over $[5/\alpha,7/\alpha]$ for $\alpha=10^{-3}$ (see figure \ref{fig:hovmoller_a1_lx07_beta0_N256}) and over $[6/\alpha,8/\alpha]$ for $\alpha=5.10^{-4}$ (see figure \ref{fig:hovmoller_a05_lx07_beta0_N256}). In both cases, the snapshot is taken at $t=6/\alpha$. We see that in both cases, the instantaneous zonal velocity profile is quite different from its temporal mean, showing that fluctuations of $U(y,t)$ are crucial in the dynamics of this flow. The parameters used in these simulations are detailed in section \ref{sub:Numerical_simulations}.}
\end{figure}

Beside its average (in time) state, it is also clear that the jet undergoes fluctuations in its position, shape and amplitude on a much shorter time-scale. In figure \ref{fig:meanU_Usnapshot_lx07_beta0_N256}, we report the comparison between snapshots of the (zonally averaged) velocity profile in the zonal direction with respect to its mean. This plot clarify more precisely that strong fluctuations are at play; it is the scope of this paper to characterise them. 

\subsection{Decomposition into zonal and non--zonal components}
Introducing non-dimensional variables in section \ref{sub:energy-balance}, the stochastic 2D Euler equations (\ref{eq:2D-NS}) turned out to depend on a parameter $\alpha$ that we defined in eq. (\ref{eq:alpha}). As already discussed, one expects this parameter to be the ratio between the time scale for the evolution of small scales $\tau$ with the time scale of evolution of large-scale coherent structures, given by the dissipative time scale $1/\lambda$. Thus, in the regime $\nu_n\ll\alpha\ll1$ we expect to observe a time scale separation between the slow evolution of large-scale jets and the fast evolution of small-scale turbulence. It is this time-scale separation that permits to find an effective dynamics of the large scales tracing out turbulent fluctuations. This has been the topic of a recent work by us  \citep{Bouchet_Nardini_Tangarife_2013_Kinetic} that we summarise in this and in the following subsections. The interested reader can consult \citep{Bouchet_Nardini_Tangarife_2013_Kinetic,nardini2014fluiddyn} for more details.

The first step is to separate the slowly evolving from the fast-evolving degrees of freedom. We thus introduce the zonal projection of a field $f$ 
\[
\left\langle f\right\rangle (y)\equiv\frac{1}{2\pi/r}\int_{0}^{2\pi/ r}f(x,y)\,\mbox{d}x.
\]
and, clearly we have $U(y)\equiv\left\langle u\right\rangle (y)$, see eq. \eqref{eq:U-new}.
Assuming that the velocity of perturbations to the zonal flow is of the order of the stochastic forcing in (\ref{eq:2D-NS}), we
decompose the velocity field as ${\bf v}=U{\bf e}_{x}+\sqrt{\alpha}{\bf v}_m$ with ${\bf v}_m=(u_m,v_m)$
and the vorticity field as $\omega=\omega_z+\sqrt{\alpha}\omega_m$ where $\omega_z\equiv\langle\omega\rangle$.
Proving that such an hypothesis is self-consistent is the most difficult part in building the kinetic theory, see sec. \ref{sub:kin-th}.

The second step is to project the 2D Euler equation \eqref{eq:2D-NS} into zonal
\begin{equation}
\frac{\partial\omega_{z}}{\partial t}=-\alpha \frac{\partial}{\partial y} \langle v_m\omega_m\rangle -\alpha\omega_{z}-\nu_{n}\left(-\partial_y^2\right)^{n}\omega_{z}
\label{eq:omega_z}
\end{equation}
and non-zonal parts,
\begin{equation}
 \frac{\partial\omega_m}{\partial t}=-L_{U}\left[\omega_m\right]-\sqrt{\alpha}NL\left[\omega_m\right]+\sqrt{2}\eta_m\,,
\label{eq:omega_m}
\end{equation}
where we have used that $C_z=0$ (i.e. the forcing does not act directly on the jet)
 and $\eta_{m}=\eta-\left\langle \eta\right\rangle $
is a Gaussian field with correlation function $\mathbb{E}\left[\eta_{m}({\bf r}_{1},t_{1})\eta_{m}({\bf r}_{2},t_{2})\right]=C_{m}({\bf r}_{1}-{\bf r}_{2})\delta(t_{1}-t_{2})$
with $C_{m}=C-\left\langle C\right\rangle $. 
In eq. \eqref{eq:omega_m}, the linear operator $L_{U}$ is
\begin{equation}
L_{U}\left[\omega_{m}\right]=L_U^0\left[\omega_{m}\right]+\alpha\omega_{m}+\nu_n\left(-\Delta\right)^n\omega_{m}\,,
\label{eq:Linearized-Dynamics}
\end{equation}
with the operator for the inertial linearised dynamics
\begin{equation}
L_{U}^0\left[\omega_{m}\right]=U(y)\frac{\partial\omega_{m}}{\partial x}-U''(y)\frac{\partial}{\partial x}\Delta^{-1}\omega_m\,,
\label{eq:Linearized-Dynamics-Inertial}
\end{equation}
where now the prime denotes the derivative with respect to $y$. Finally, the non-linear operator $NL$ reads
\[
NL\left[\omega_m\right]=  v_m\omega_m -\langle v_m\omega_m\rangle.
\]
In the following, instead of dealing with the equation for $\omega_z$, see eq. (\ref{eq:omega_z}), 
it will be more convenient to consider the equation for the average velocity profile $U(y)$, which can be obtained remembering that  $\omega_z(y) = -U'(y)$. One has
\begin{equation}
\frac{\partial U}{\partial t}=\alpha  \langle v_m\omega_m\rangle -\alpha U-\nu_{n}\left(-\partial_y^2\right)^{n}U\,.
\label{eq:U}
\end{equation}
Using the Taylor relation $\langle v_m\omega_m\rangle=\partial_y \langle v_m u_m\rangle$, we see that $\langle v_m\omega_m\rangle$ is the divergence of a Reynolds stress component. In the following, $\langle v_m\omega_m\rangle$ will be called the Reynolds stress divergence.\\
The third and final step to obtain our kinetic equation for the slow evolution of $U(y,t)$ is summarised in the next paragraph.

\subsection{Kinetic equation for the slow evolution of jets}\label{sub:kin-th}
From \eqref{eq:omega_m} and \eqref{eq:U} it is possible to derive an effective evolution for the zonal velocity profile $U(y,t)$, valid in the regime $\nu_n\ll\alpha\ll1$. The approach has been developed in \citep{Bouchet_Nardini_Tangarife_2013_Kinetic,nardini2014fluiddyn} and has been named kinetic theory because it shares strong similarities with the kinetic theories developed for plasma and gravitational systems \citep{Nicholson_1991,balescu1975equilibrium,Nardini_Gupta_Ruffo_Dauxois_Bouchet_2012_kinetic,NardiniGuptaBouchet-2012-JSMTE,heyvaerts2010balescu}. It could have been also named   stochastic averaging, borrowing from mathematics the name of the technique employed to derive the effective evolution \citep{Gardiner_1994_Book_Stochastic,pavliotis2008multiscale}.

Kinetic theory provides the effective slow dynamics of $U(y,t)$ with the fast evolution of $\omega_m({\bf r},t)$ adiabatically relaxed to its statistically stationary state with a fixed background flow $U(y)$. Such effective dynamics can be obtained with a perturbative expansion in $\alpha\ll1$ (stochastic averaging, see \citep{Gardiner_1994_Book_Stochastic,pavliotis2008multiscale}). In the case of the stochastic 2D Euler equations (\ref{eq:omega_m},\ref{eq:U}), the resulting effective slow dynamics reads \citep{Bouchet_Nardini_Tangarife_2013_Kinetic,nardini2014fluiddyn}
\begin{equation}
\frac{\partial U}{\partial t} = \alpha F^0[U]  - \alpha U + \alpha \xi[U]\,.
\label{eq:kinetic_equation}
\end{equation}
where
\begin{equation}
F^0[U](y) = \mathbb{E}_U^0\langle v_m \omega_m\rangle(y)
\label{eq:F_U_def}
\end{equation}
and $\xi[U]$ is a Gaussian noise with zero mean and correlations (for a fixed $U(y)$)
\begin{equation}
\mathbb{E}\left[\xi[U](y_1,t_1) \xi[U](y_2,t_2)\right] = \delta(t_1-t_2) \Xi^0[U](y_1,y_2)
\end{equation}
with
\begin{equation}
\Xi^0[U](y_1,y_2) = \int_0^\infty \mathbb{E}_U^0\left[\left[\,\langle v_m \omega_m\rangle(y_1,s)\langle v_m \omega_m\rangle(y_2,0) + (y_1\leftrightarrow y_2)\,\right]\right]\,\mathrm{d}s\,,
\label{eq:Xi_U_def}
\end{equation}
where $(y_1\leftrightarrow y_2)$ denotes the symmetric expression obtained inverting $y_1$ and $y_2$. In \eqref{eq:F_U_def}, the operator $\mathbb{E}_U^0[\cdot]$ denotes the expectation over the stationary distribution of the effective fast dynamics of $\omega_m$,
\begin{equation}
 \frac{\partial\omega_m}{\partial t}=-L_{U}^0\left[\omega_m\right]+\sqrt{2}\eta_m\,,
\label{eq:omega_m_linear_inertial}
\end{equation}
where $U(y)$ is fixed and $L_U^0$ is given by \eqref{eq:Linearized-Dynamics-Inertial}. In \eqref{eq:Xi_U_def}, the operator $\mathbb{E}_U^0[[\cdot]]$ denotes the covariance over the stationary distribution of \eqref{eq:omega_m_linear_inertial},
\begin{equation}
\mathbb{E}_U^0\left[\left[\,f[\omega_m]g[\omega_m]\,\right]\right] \equiv \mathbb{E}_U^0\left[f[\omega_m]g[\omega_m]\right] - \mathbb{E}_U^0\left[f[\omega_m]\right]\mathbb{E}_U^0\left[g[\omega_m]\right]\,.
\label{eq:covariance_def}
\end{equation}
Using that $\mathbb{E}_U^0$ is an average in the statistically stationary state of \eqref{eq:omega_m_linear_inertial}, we easily get
\begin{equation}
\Xi^0[U](y_1,y_2) = \lim_{\Delta t\to\infty} \mathbb{E}_U^0\left[\left[ \frac{1}{\Delta t}\int_0^{\Delta t}\mathrm{d}s_1\int_0^{\Delta t}\mathrm{d}s_2\,\langle v_m \omega_m\rangle(y_1,s_1)\langle v_m \omega_m\rangle(y_2,s_2) \,\right]\right]\,,
\end{equation}
i.e. $\Xi^0[U](y_1,y_2)$ is the covariance of the time-averaged Reynolds stress divergence, properly rescaled in the limit of infinite time-averaging window $\Delta t\to\infty$. In other words, $\Xi^0[U]$ contains the information about the Gaussian statistics of time-averaged Reynolds stresses, corresponding to the Central Limit Theorem \citep{bouchet2015large,freidlin2012random}.

Note that because we are investigating the regime $\nu_n\ll\alpha$ and because (hyper-)viscosity acts predominantly at small scales, viscous dissipation is negligible in the effective dynamics \eqref{eq:kinetic_equation}. Besides, non-linear terms $NL$ in \eqref{eq:omega_m} are of order $\sqrt\alpha$, which explains why they do not appear in the leading order description \eqref{eq:omega_m_linear_inertial}. Then, \eqref{eq:omega_m_linear_inertial} is a linear equation forced by a Gaussian white noise (Ornstein-Uhlenbeck process \citep{Gardiner_1994_Book_Stochastic}). This property will be crucial in our analysis.

Similar effective descriptions to our \eqref{eq:kinetic_equation} 
were obtained previously in phenomenological ways by using either a quasi-linear approximation of the dynamics (i.e. neglecting the term $NL[\omega_m]$ in \eqref{eq:omega_m} \citep{Srinivasan-Young-2011-JAS}) or using a closure in the hierarchy for the cumulants of the vorticity \citep{ait2015cumulant}. Such approaches have been called Stochastic Structural Stability Theory (S3T \citep{bakas2015s3t,farrell2003structural}) or Cumulant Expansion at Second order (CE2 \citep{marston2010statistics,Srinivasan-Young-2011-JAS,tobias2013direct}). \\
We note, however, that i) phenomenological approaches were not able to capture the precise form of the average $\mathbb{E}_U^0$ entering in the kinetic equation and, more importantly, ii) they only captured the deterministic part of the kinetic equation \eqref{eq:kinetic_equation} (without the noise term $\xi[U]$).
By contrast, the kinetic equation \eqref{eq:kinetic_equation} arises from a formal perturbative expansion in powers of $\alpha\ll1$, and thus justifies the quasi-linear approximation (or equivalently the closure in the hierarchy for cumulants) in this regime \citep{Bouchet_Nardini_Tangarife_2013_Kinetic}. Moreover, kinetic theory goes beyond S3T-CE2 approaches as it us expected to describe fluctuations of jets around their attractors, through the noise term $\xi[U]$ in \eqref{eq:kinetic_equation}.\\

The main achievement of our previous work \citep{Bouchet_Nardini_Tangarife_2013_Kinetic} has been to prove that the average Reynolds stress divergence $F^0[U]$ is finite, i.e. ultraviolet divergences are not present in our perturbative approach when considering the kinetic equation up to order $\mathcal{O}(\alpha)$. This result is striking, as we are dealing with the dynamics \eqref{eq:omega_m_linear_inertial} with stochastic forcing but no energy dissipation and no viscous regularization at small scales. The properties of $F^0[U]$ are a consequence of inviscid damping mechanisms, known for the linearized 2D Euler dynamics as the Orr mechanism and the depletion of vorticity at the stationary streamline \citep{Bouchet_Morita_2010PhyD,Orr_1907}, that will be reviewed in section \ref{sub:Orr}.

The goal of this paper is to study the effect of the noise term $\xi[U]$ on the evolution of the zonal velocity profile.
Formally, $\Xi^0[U]$ is defined as the infinite-time limit of an expectation of the process \eqref{eq:omega_m_linear_inertial}, where no dissipation is present. For convenience, we will also consider the linear dynamics of $\omega_m$ with a small but non-zero friction coefficient $\alpha$,
\begin{equation}
 \frac{\partial\omega_m}{\partial t}=-L_{U}^0\left[\omega_m\right] - \alpha \omega_m+\sqrt{2}\eta_m\,,
\label{eq:omega_m_linear}
\end{equation}
the expectation in the statistically stationary state of \eqref{eq:omega_m_linear} will be denoted $\mathbb{E}^\alpha_U$. Then we define
\begin{equation}
\Xi^\alpha[U](y_1,y_2) = \int_0^\infty \mathbb{E}_U^\alpha\left[\left[\,\langle v_m \omega_m\rangle(y_1,s)\langle v_m \omega_m\rangle(y_2,0) + (y_1\leftrightarrow y_2)\,\right]\right]\,\mathrm{d}s\,,
\label{eq:Xi_U_alpha_def}
\end{equation}
and we will be interested in the limit for $\alpha\to0$ of $\Xi^\alpha[U]$. 


\section{Gaussian fluctuations of Reynolds stresses in the inertial limit\label{sec:Fluctuations_Reynolds}}

We derive in this section our main theoretical result, expressed in eq. \eqref{eq:SC-Xi-Xi_kl} and \eqref{eq:SC-Xi-kl-convergence}: these equations give the spatial structure of Reynolds stresses covariance $\Xi^\alpha[U](y_1,y_2)$, at leading order when $\alpha\ll1$ and for $\nu_n=0$. Then in section \ref{sec:Consequences_mean_flow} we discuss the implications of eq. \eqref{eq:SC-Xi-Xi_kl} and \eqref{eq:SC-Xi-kl-convergence} on the statistics of fluctuations of $U(y,t)$ and compare them to results from direct numerical simulations. If we were dealing with a system with a finite number of degrees of freedom, one would expect the noise $\alpha \xi$ in the kinetic equation \eqref{eq:kinetic_equation} to give a contribution only at order $\mathcal{O}(\alpha^2)$. However, we show in this section that ultraviolet divergences renormalise such result, leading to eq. \eqref{eq:SC-Xi-Xi_kl} and \eqref{eq:SC-Xi-kl-convergence}.

We start our derivation with some definitions in paragraph \ref{sec:IM-Fourier}. 
 In paragraph \ref{sub:Orr}, we discuss the technical results (Orr mechanism) that permits to derive \eqref{eq:SC-Xi-kl-convergence} in the general case. Then, in paragraph \ref{sub:autocorr-linear-shear} we derive \eqref{eq:SC-Xi-kl-convergence} in a simple esplicitely solvable case, when the background flow $U(y)$ is a constant shear. The proof of \eqref{eq:SC-Xi-kl-convergence} for a general background flow, being rather technical, is left  in Appendix \ref{appendix:proof}. However, the main technical points to obtain our result will be transparent to the reader after he has read the present section.

\subsection{Fourier decomposition and autocorrelation function in terms of two-points correlations \label{sec:IM-Fourier}}
Because the dynamics of non-zonal  vorticity $\omega_m$ is linear, see eq.  \eqref{eq:omega_m_linear}, the dynamics of each Fourier mode can be studied independently and the global result will be obtained by simply adding the contribution from each mode. We treat here the simple case of a flow in a biperiodic domain $\mathcal{D} = [0,2\pi /r)\times[0,2\pi r)$, the generalization to different geometries being straightforward. In the domain $\mathcal{D}$, the wavevectors read ${\bf k} = (k,l)$ with $k/r\in\mathbb{Z}$ and $l\times r\in\mathbb{Z}$.\\
We begin expanding the force correlation function $C_{m}$ in Fourier series,
\begin{equation}
C_{m}(x,y)=\sum_{k>0\,,l}c_{kl}\cos(kx+ly),\label{eq:SC-Cm}
\end{equation}
with $c_{kl}\geq0$. We note that because $C_{m}$ is a correlation, it is a positive definite
function, implying the absence of $\sin$ contributions in the above
expansion. 
The generalization to the case of an inhomogeneous force, for instance
for the case of a channel is straightforward. The noise correlation function $C_m$ corresponds to the noise
\begin{equation}
\eta_{m}(\mathbf{r},t)=\sum_{k=-\infty}^{\infty}\sum_{l=-\infty}^{\infty}\sqrt{\frac{c_{kl}}{2}}\,\mbox{e}^{ikx+ily}\eta_{kl}(t)
\label{eq:IM-Fourier-eta_m}
\end{equation}
where $\eta_{kl}^{*}=\eta_{-k,-l}$ and $\mathbb{E}[ \eta_{k_{1},l_{1}}(t_1)\eta_{k_{2},l_{2}}(t_2)]=\delta_{k_{1},-k_{2}}\delta_{l_{1},-l_{2}}\delta(t_1-t_2)$,
and $c_{k,l}$ is defined for $k<0$ by $c_{k,l}=c_{-k,-l}$, and for $k=0$ by $c_{0,l}=0$. In the following, all sums over ${\bf k} = (k,l)$ include negative $k$.

Let us now consider the dynamics of the non-zonal vorticity in eq. \eqref{eq:omega_m_linear}. Because $L_{U}$ is linear and invariant under translations in the $x$ direction, the non-zonal vorticity field can be written as 
\begin{equation}
\omega_m(\mathbf{r},t) = \sum_{{\bf k}}\sqrt{\frac{c_{kl}}{2}}\,\mathrm{e}^{ikx}\omega_{kl}(y,t)\,,
\label{eq:SC-omega_kl}
\end{equation}
where $\omega_{kl}$ evolves according to
\begin{equation}\label{eq:linear-qkl}
\frac{\partial\omega_{kl}}{\partial t}+L_{U,k}^0[\omega_{kl}]=-\alpha\omega_{kl}+\sqrt2\mbox{e}^{ily}\eta_{kl}\,,
\end{equation}
where
\begin{equation}
L_{U,k}^0[\omega_{kl}]=ikU(y)\omega_{kl} - ikU''(y)\Delta_k^{-1}\omega_{kl}
\end{equation}
with $\Delta_k=\partial_y^2- k^2$.

Let us now consider the decomposition of $\Xi^\alpha[U]$ into Fourier modes. Using \eqref{eq:Xi_U_def} and \eqref{eq:SC-omega_kl}, we get
\[\begin{aligned}
\Xi^\alpha[U](y_1,y_2)\! = \!\sum_{{\bf k}, {\bf k}'} \frac{c_{kl}}{2}  \frac{c_{k'l'}}{2}\!\int_0^\infty \!\mathbb{E}_U^\alpha&\left[\left[ \, \left(v_{kl}\omega_{-k,-l}\right) (y_1,s) \left(v_{k'l'}\omega_{-k',-l'}\right) (y_2,0)+ (y_1\leftrightarrow y_2) \,\right]\right]\,\mathrm{d} s,
\end{aligned}\]
where we recall that $\mathbb{E}_U^\alpha[[\cdot]]$ denotes the covariance in the stationary state of \eqref{eq:linear-qkl}. In order to analyse the above expression is very useful to remember that, as $\omega_{kl}$ (defined in \eqref{eq:linear-qkl}) obey to an Ornstein-Uhlenbeck processes with zero initial condition, so they are Gaussian random variables at all times \citep{Gardiner_1994_Book_Stochastic}. Moreover, $v_{kl}$ are obtained via a linear transformation of $\omega_{kl}$ so that they are also Gaussian random variables at all times. The Isserlis-Wick theorem can then be applied so that we can reduce the four-points correlation functions in products of two-points correlation functions. Moreover, using the fact that $\omega_{k_{1},l_{1}}$ and $\omega_{k_{2},l_{2}}^*$ are statistically independent for $(k_1,l_1)\neq(k_2,l_2)$, we get
\begin{equation}
\Xi^\alpha[U](y_1,y_2)  =\sum_{{\bf k}} c_{kl}^2\, \big\lbrace\Xi_{kl}^\alpha(y_1,y_2)+\Xi_{kl}^\alpha(y_2,y_1)\big\rbrace
\label{eq:SC-Xi-Xi_kl}
\end{equation}
with $\Xi_{kl}^\alpha(y_1,y_2)=C_{kl}^\alpha(y_{1},y_{2})+D_{kl}^\alpha(y_{1},y_{2})$ where
\begin{equation}
C_{kl}^\alpha(y_{1},y_{2})=\frac14\int_{0}^\infty \mathbb{E}_U^\alpha\left[  v_{kl}(y_1,s) v_{-k,-l}(y_2,0) \right]\mathbb{E}_U^\alpha\left[  \omega_{-k,-l} (y_1,s) \omega_{kl}(y_2,0) \right]\,\mathrm{d} s
\label{eq:C-vv}
\end{equation}
and
\begin{equation}
D_{kl}^\alpha(y_{1},y_{2})=\frac14\int_{0}^\infty\mathbb{E}_U^\alpha\left[  v_{kl}(y_1,s) \omega_{-k,-l}(y_2,0) \right]\mathbb{E}_U^\alpha\left[   \omega_{-k,-l} (y_1,s) v_{kl}(y_2,0) \right]\,\mathrm{d} s.\label{eq:C-vq}
\end{equation}
Note that, by definition of the covariance \eqref{eq:covariance_def}, the two-point correlations involving $\omega$ and $v$ evaluated at the same spatial point have been cancelled in the computation of $\Xi_{kl}^\alpha$.

Following a classical procedure for the computation of two-points correlation functions of Ornstein-Uhlenbeck processes \citep{Bouchet_Nardini_Tangarife_2013_Kinetic,Gardiner_1994_Book_Stochastic}, the two-points correlation functions appearing in \eqref{eq:C-vv} and \eqref{eq:C-vq} can be expressed as 
\begin{equation}
T^\alpha_{\omega\omega}(k,l,y_{1},y_{2},s)\equiv
\frac12\mathbb{E}_U^\alpha\left[\omega_{-k,-l}(y_{1},s)\omega_{k,l}(y_{2},0)\right]=
\int_{0}^{\infty}\mbox{d}t_1\,\tilde{\omega}_{-k,-l}(y_{1},s+t_1)\tilde{\omega}_{k,l}(y_{2},t_1)\,,\label{eq:convergence-CR-qq}
\end{equation}
\begin{equation}
T^\alpha_{vv}(k,l,y_{1},y_{2},s)\equiv
\frac12\mathbb{E}_U^\alpha\left[v_{k,l}(y_{1},s)v_{-k,-l}(y_{2},0)\right]=
 \int_{0}^{\infty}\mbox{d}t_1\,\tilde{v}_{k,l}(y_{1},s+t_1)\tilde{v}_{-k,-l}(y_{2},t_1)\,,\label{eq:convergence-CR-vv}
\end{equation}
\begin{equation}
T^\alpha_{v\omega}(k,l,y_{1},y_{2},s)\equiv
\frac12\mathbb{E}_U^\alpha\left[v_{k,l}(y_{1},s)\omega_{-k,-l}(y_{2},0)\right]=
\int_{0}^{\infty}\mbox{d}t_1\,\tilde{v}_{k,l}(y_{1},s+t_1)\tilde{\omega}_{-k,-l}(y_{2},t_1)\, ,\label{eq:convergence-CR-vq}
\end{equation}
\begin{equation}
T^\alpha_{\omega v}(k,l,y_{1},y_{2},s)\equiv
\frac12\mathbb{E}_U^\alpha\left[\omega_{-k,-l}(y_{1},s)v_{k,l}(y_{2},0)\right]=
\int_{0}^{\infty}\mbox{d}t_1\,\tilde{\omega}_{-k,-l}(y_{1},s+t_1)\tilde{v}_{k,l}(y_{2},t_1)\, ,\label{eq:convergence-CR-qv}
\end{equation}
where $\tilde{\omega}_{kl}$
is the solution of the deterministic linear dynamics
\qq\label{eq:eddy-vorticity-linear-fourier}
\partial_{t}\tilde{\omega}_{kl}+L_{U,k}^0[\tilde{\omega}_{kl}]+\alpha \tilde{\omega}_{kl} =0\qquad \qquad \textrm{with initial condition } \qquad \tilde{\omega}_{kl}(y,0)=\mbox{e}^{ily}
\qqq
and $\tilde{v}_{kl}$ is the associated meridional velocity. Equations (\ref{eq:convergence-CR-qq}--\ref{eq:convergence-CR-qv}) give two-points correlation functions in terms of time integrals of deterministic fields. The properties of these correlation functions, and of $\Xi^\alpha[U](y_1,y_2)$, thus depend on the asymptotic behaviour of these deterministic fields. We now describe in details this asymptotic behaviour.

\subsection{Inviscid damping mechanism for the deterministic linear Euler dynamics\label{sub:Orr}}

\begin{figure}\begin{center}
\includegraphics[height=3cm]{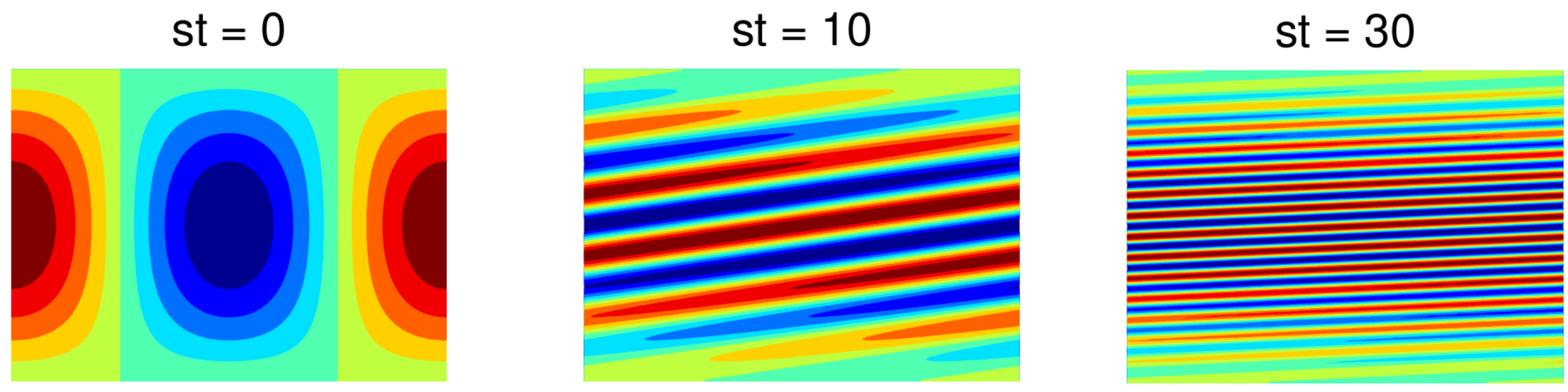}
\caption{Evolution of the perturbation vorticity, advected by the constant shear base flow $U(y)=\sigma y$.\label{figure-orr}}
\end{center}\end{figure}

We have seen in the previous paragraph that two-points correlation functions of the Ornstein-Uhlenbeck process $\omega_m$ can be computed from time-integrals (\ref{eq:convergence-CR-qq}--\ref{eq:convergence-CR-qv}), involving solutions of the associated deterministic problem in eq. (\ref{eq:eddy-vorticity-linear-fourier}). When $\alpha\neq0$, the deterministic vorticity field $\tilde{\omega}_{kl}(y,t_1)$ decays exponentially with rate $\alpha$ in the limit $t_1\to\infty$. Then, time integrals (\ref{eq:convergence-CR-qq}--\ref{eq:convergence-CR-qv}) always converge. Moreover, the time-correlation functions given by (\ref{eq:convergence-CR-qq}--\ref{eq:convergence-CR-qv}) also decay exponentially with rate $\alpha$ for large $s$, so that integrals (\ref{eq:C-vv},\ref{eq:C-vq}) always converge, and $\Xi^\alpha[U]$ is always finite for $\alpha\neq0$.

However, we are specifically interested in the regime $\alpha\to0$. Indeed, by definition the noise correlation $\Xi^0[U]$ in \eqref{eq:Xi_U_def} is defined for the linear dynamics \eqref{eq:omega_m_linear_inertial} with no friction. As a consequence, the convergence of integrals (\ref{eq:convergence-CR-qq}--\ref{eq:convergence-CR-qv}) should rely only on an inviscid damping mechanism of the inertial deterministic linear dynamics $\partial_{t}+L_{U,k}^0$. This inviscid damping is known for the linearized 2D Euler dynamics as the Orr mechanism and the depletion of vorticity at the stationary streamline \citep{Bouchet_Morita_2010PhyD,Orr_1907}\\

The phenomenology is the following: while the vorticity shows filaments at finer and finer scales when
time increases, non-local averages of the vorticity (such as the one leading to the computation of the streamfunction or the velocity) converge to zero in the long time limit.

As an example, consider the case of the linear Euler equation in a channel $(x,y)\in\mathcal{D}=[0,2\pi L_x)\times[0,L_y]$, or in an infinite domain $(x,y)\in\mathcal{D}=[0,2\pi L_x)\times\mathbb{R}$, where the background flow is $U(y)=\sigma y$ with a constant shear rate $\sigma$. Then $U''=0$ and $L_{U,k}^0 = ik\sigma y $. This is actually the case first studied by Orr \citep{Orr_1907}. According to the discussion of the previous paragraph, we consider the deterministic linear dynamics
\begin{equation}
\frac{\partial \tilde{\omega}_{k,l}}{\partial t} +ik\sigma y\,\tilde{\omega}_{k,l}(y,t)=0\quad,\quad \tilde{\omega}_{k,l}(y,0)=\mbox{e}^{ily},
\end{equation}
which can be solved as $\tilde{\omega}_{k,l}(y,t)=\mbox{e}^{-ik\sigma yt+ily}$. This increasing filamentation of the vorticity field as time goes on can be seen in figure \ref{figure-orr}. The meridional velocity is then computed as
\begin{equation}
\tilde{v}_{kl}(y,t) = ik\int \md y' \, H_k(y,y')\tilde{\omega}_{k,l}(y',t)\,,
\end{equation}
where $H_k$ is the Green function of the Laplacian $\Delta_k = \partial_y^2 -k ^2$, i.e. such that $\Delta _k H_k(y,y')  = \delta(y-y')$. Such integral is an oscillating integral. In the limit $t\to\infty$, it decays algebraically to zero with a power that depends on the order of differentiability of $H_k$. In this case, Orr proved \citep{Orr_1907} that 
\begin{equation}
\tilde{v}_{kl}(y,t) \underset{t\to\infty}{\sim} \frac{\mbox{e}^{-ik\sigma yt+ily}}{ik\sigma ^{2}\,t^{2}}\,.\label{eq:IM-Orr-stream}
\end{equation}
The filamentation and the related relaxation mechanism with no dissipation for the velocity and streamfunction is very general for advection equations and it has an analog in plasma physics in the context of the Vlasov
equation, where it is called Landau damping \citep{Nicholson_1991,villani2010landau}.

We note that in \eqref{eq:IM-Orr-stream}, the shear $\sigma $ plays the role of an effective damping rate. The generalization of the Orr mechanism to the case of any strictly monotonic profile $U(y)$ ---i.e. when the shear is always non-zero--- has been first considered~\citep{brown1980algebraic}. However, zonal jets necessarily have velocity extrema. The generalization of the Orr mechanism to non-monotonic background flows $U(y)$ has only been considered recently \citep{Bouchet_Morita_2010PhyD}. Under the assumption that the linear operator $L_{U,k}^{0}$ has no modes, it has been shown that~\citep{Bouchet_Morita_2010PhyD}
\begin{equation}
\tilde{\omega}_{kl}(y,t)\underset{t\to\infty}{\sim}\tilde{\omega}_{kl}^{\infty}(y)\mbox{e}^{-ikU(y)t}\,,\label{eq:orr-mechanism-voricity}
\end{equation}
where the function $\tilde{\omega}_{kl}^{\infty}(y)$ depends on the whole velocity profile $U(y)$. The Orr mechanism for $U(y)=\sigma y$ is a particular case of \eqref{eq:orr-mechanism-voricity}, where $\tilde{\omega}_{kl}^\infty(y)=\me^{ily}$. Using again results on oscillating integrals and the properties of the Laplacian Green function $H_k$, we have the asymptotic decay of the meridional velocity \citep{Bouchet_Morita_2010PhyD}
\begin{equation}
\tilde{v}_{kl}(y,t)\underset{t\to\infty}{\sim}\frac{\tilde{\omega}_{kl}^{\infty}(y)}{ik(U'(y))^{2}}\frac{\mbox{e}^{-ikU(y)t}}{t^{2}}.\label{eq:orr-mechanism-velocity-y}
\end{equation}
In (\ref{eq:orr-mechanism-voricity}) and (\ref{eq:orr-mechanism-velocity-y}), higher order corrections are present and
decay with higher powers in $1/t$. 

Mathematical proofs of the asymptotic behaviour (\ref{eq:orr-mechanism-voricity}), (\ref{eq:orr-mechanism-velocity-y}) have been given recently, either for the case of a strictly monotic profile $U(y)$ \citep{zillinger2014linear,zillinger2015linear} or for the relaxation of the non-linear 2D Euler equation after a small perturbation of the constant shear profile $U(y)=\sigma y$ \citep{bedrossian2013inviscid}, following the analogous theorem for non-linear Landau damping \citep{mouhot2011landau}.

At this stage, a natural question is: what happens when the local shear vanishes? Indeed, a jet profile necessarily presents extrema of the velocity, at points $y_0$ such that $U'(y_0)=0$. Such points are called stationary points of the zonal jet profile. It can be shown that at the stationary points, the perturbation vorticity also decays for large times: $\tilde{\omega}_{kl}^\infty(y_0)=0$ \citep{Bouchet_Morita_2010PhyD}. This phenomenon has been called vorticity depletion at the stationary
streamlines. It has been observed numerically that the extend of the
area for which $\tilde{\omega}_{kl}^{\infty}(y_{0})\simeq0$ can be
very large, up to half of the total domain, meaning that in a large
part of the domain, the shear is not the explanation for the asymptotic
decay. The formula for the vorticity (\ref{eq:orr-mechanism-voricity})
is valid for any $y$. The formulas for the velocity and stream functions
are valid for any $y\neq y_{0}$. Exactly at the specific point $y=y_{0}$,
the damping is still algebraic with preliminary explanation given
in \citep{Bouchet_Morita_2010PhyD}, but a complete theoretical prediction
is not yet available.

Equations \eqref{eq:orr-mechanism-voricity} and \eqref{eq:orr-mechanism-velocity-y} give the asymptotic behaviour of vorticity and meridional velocity in the deterministic linear 2D Euler equation, with no external damping mechanism. In the following, we will also be interested in the behaviour of these fields when a small friction or viscosity are present. For simplicity, we will only treat the case of a small friction (which acts uniformly at all scales): $\nu=0$. Then, the linear friction leads to an exponential damping of all fields, with rate $\alpha$. It will be useful to generalize \eqref{eq:orr-mechanism-voricity} as
\begin{equation}
\tilde{\omega}_{kl}^\alpha(y,t)\underset{}{=}\tilde{\omega}_{kl}^{\infty}(y)\mbox{e}^{-(ikU(y)+\alpha)t}\,+\tilde{\omega}_{kl}^{r,\alpha}(y,t)\,.
\label{eq:orr-mechanism-voricity-alpha}
\end{equation}
The above formula defines
$\tilde{\omega}_{kl}^{r,\alpha}$. The classical Orr mechanism \eqref{eq:orr-mechanism-voricity} is equivalent to the statement that for all values of $\alpha$ (even for $\alpha=0$), $\tilde{\omega}_{kl}^{r,\alpha}(y,t)$ is a bounded function both in $y$ and $t$, and decays to 0 as $t\to\infty$. Actually, a refined formulation of the Orr mechanism is that $\tilde{\omega}_{kl}^{r,\alpha}(y,t)\underset{t\to\infty}{\sim}O(\me^{-\alpha t}/t^{\gamma})$, with $\gamma>0$~\citep{Bouchet_Morita_2010PhyD}.\\

We have thus seen that, under the hypothesis that the linear operator $L_{U,k}^0$ has no modes, the deterministic linear dynamics of the eddies leads to an inviscid damping of the velocity and of the streamfunction. These results form furnish the basis to prove our central result of this paper that will be explained in the next section.

\subsection{Integrated autocorrelation function in the inertial limit}\label{sub:autocorr-linear-shear}
The central theoretical result of the paper is that, for $\nu_n=0$ and in the limit of $\alpha\to0$, the (spectral content of the) covariance of the noise entering in the kinetic equation (\ref{eq:kinetic_equation}) is given by
\begin{equation}
\Xi^\alpha_{kl}(y_1,y_2)\;\underset{\alpha\to0}{\sim}\; \frac{A_{kl}(y_1,y_2)}{ik(U(y_1)-U(y_2))+2\alpha},
\label{eq:SC-Xi-kl-convergence}
\end{equation}
where $A_{kl}$ is a regular function independent of $\alpha$. The full covariance can be then obtained by summing up different Fourier modes, see eq. (\ref{eq:SC-Xi-Xi_kl}).
This result will permit us to predict in section \ref{sec:covariance-U} the statistics of the Gaussian fluctuations of $U(y)$.

Before entering in the derivation of eq. (\ref{eq:SC-Xi-kl-convergence}), few comments are mandatory. First,  at points such that $U(y_1)=U(y_2)$, we readily see that $\Xi^\alpha_{kl}(y_1,y_2)$ behaves like $1/\alpha$. Using Plemelj formula
\begin{equation}
\lim_{\alpha\rightarrow0^{+}}\frac{-i}{y-i\alpha}=\pi\delta\left(y\right)-iPV\left(\frac{1}{y}\right),\label{eq:Plemelj}
\end{equation}
where $PV$ is the Cauchy Principal Value distribution, we also see that $\Xi^\alpha_{kl}(y_1,y_2)$ converges in the sense of distributions as $\alpha\to0$. Secondly, we show in Appendix \ref{appendix:proof}
 that at points such that $U(y_1)=U(y_2)$ and if $U'(y_1)=0$, then $A_{kl}(y_1,y_2)=0$. This means that, at such points, $\Xi^\alpha_{kl}(y_1,y_2)$ either converges to a finite value or diverges slower than $1/\alpha$ as $\alpha\to0$. Finally, as the total covariance of the noise $\Xi^\alpha[U](y_1,y_2)$ is obtained by a linear superposition of different modes contributions, see \eqref{eq:SC-Xi-Xi_kl}, the very same behaviour is expected for $\Xi^\alpha[U](y_1,y_2)$.\\

In this section we only prove eq. (\ref{eq:SC-Xi-kl-convergence}) in the simple case of $U(y)$ being a constant shear; this case can indeed be handled easily base because the deterministic linear equation for the eddy vorticity (\ref{eq:eddy-vorticity-linear-fourier}) can be solved analytically. The general proof for any background flow $U(y)$, being rather technical, is left in Appendix \ref{appendix:proof}: it is based on estimating the large $s$-behaviour of $T^\alpha_{\omega\omega}$, $T^\alpha_{vv}$, $T^\alpha_{v\omega}$ and $T^\alpha_{\omega v}$ by using the asymptotic behaviours in eq. (\ref{eq:orr-mechanism-velocity-y}) and (\ref{eq:orr-mechanism-voricity-alpha}) as described in previous section.

\subsubsection{Explicit computation in the case of a constant shear}\label{subsub:autocorr-linear-shear}
We consider here the case of the linear Euler equation in a channel $(x,y)\in\mathcal{D}=[0,2\pi L_x)\times[0,L_y]$, or in an infinite domain $(x,y)\in\mathcal{D}=[0,2\pi L_x)\times\mathbb{R}$, where the background flow is $U(y)=sy$ with a constant shear $s$. In this case, the deterministic linear equation can be solved explicitely, and all the quantities of interest can be expressed in terms of spatial integrals involving $H_k$, the Green function of the Laplacian $\Delta_k = \partial_y^2 -k ^2$. In the following, we will not need the explicit expression of $H_k$, but only the fact that $H_k$ is a continuous function of its two variables, and that the first derivative $\partial_y H_k(y,y')$ is discontinous at $y=y'$, see \citep{Bouchet_Morita_2010PhyD}.\\

The correlation functions $T^\alpha_{vv}$ and $T^\alpha_{\omega\omega}$ can be easily computed injecting the expressions of $\tilde{\omega}_{kl}$ and $\tilde{v}_{kl}$ into \eqref{eq:convergence-CR-qq} and \eqref{eq:convergence-CR-vv}, leading to
 \begin{equation}\begin{aligned}
C^\alpha_{kl}(y_1,y_2) &\equiv \frac14 \int_0^\infty \left[T^\alpha_{\omega\omega}\cdot T^\alpha_{vv}\right](k,l,y_1,y_2,s) \,\md s \\
& =-\frac{i}{ks^{3}}\frac{\mbox{e}^{-il(y_{1}-y_{2})}}{y_{1}-y_{2}+\frac{2\alpha}{ks}i}\int \mbox{d}y'_{1}\,\int \mbox{d}y'_{2}\,\frac{H_{k}(y_{1},y'_{1})H_{-k}(y_{2},y'_{2})\mbox{e}^{il(y'_{1}-y'_{2})}}{\left(y'_{2}-y'_{1}+\frac{2\alpha}{ks}i\right)\left(y_{1}-y'_{1}+\frac{2\alpha}{ks}i\right)}\,.\\
\end{aligned}\label{eq:linear-shear-Cvv-alpha}\end{equation}
$H_{k}$ is a continuous function, so the spatial integrals appearing in the above expression converge to a finite quantity in the limit $\alpha\to0$:
\begin{equation}
\int \mbox{d}y'_{1}\,\int \mbox{d}y'_{2}\,\frac{H_{k}(y_{1},y'_{1})H_{-k}(y_{2},y'_{2})\mbox{e}^{il(y'_{1}-y'_{2})}}{\left(y'_{2}-y'_{1}+\frac{2\alpha}{ks}i\right)\left(y_{1}-y'_{1}+\frac{2\alpha}{ks}i\right)} \;\underset{\alpha\to0}{\longrightarrow}\;A(k,l,y_1,y_2)
\end{equation}
where $A$ is a regular function independent of $\alpha$, that can be written explicitely using Plemelj formula \eqref{eq:Plemelj}. Then, we clearly see that, due to the pre-factor in \eqref{eq:linear-shear-Cvv-alpha}, $C^\alpha_{kl}(y_1,y_2)$ is finite for $y_{1}\neq y_{2}$ and diverges as $1/\alpha$ for $y_{1}=y_{2}$.

Similarly, we can compute
 \begin{equation}
 \begin{aligned}
D^\alpha_{kl}(y_{1},y_{2})& \equiv \frac14 \int_0^\infty \left[T^\alpha_{v\omega}\cdot T^\alpha_{\omega v}\right](k,l,y_1,y_2,s) \,\md s \\
&=-\frac{i}{ks^{3}}\left(\int \mbox{d}y'_{1}\,
\frac{H_{k}(y_{1},y'_{1})\mbox{e}^{-il(y'_{1}-y_{2})}}{\left(y_{2}-y'_{1}+\frac{2\alpha}{ks}i\right)\left(y_{1}-y'_{1}+\frac{2\alpha}{ks}i\right)}\right)\times\\
&\qquad\qquad\left(\int \mbox{d}y'_{2}\,\frac{H_{k}(y_{2},y'_{2})\mbox{e}^{-il(y'_{2}-y_{1})}}{y_{1}-y'_{2}+\frac{2\alpha}{ks}i}\right)\,.
\label{eq:linear-shear-Cvq-alpha}
\end{aligned}\end{equation}
We observe that this expression is the product of two integrals. The second one converges to a finite quantity when $\alpha\to0$, for any $y_1$ and $y_2$. Moreover, if $y_1\neq y_2$ the first integral also has a finite limit, using again that the Green function $H_k$ is continuous. However, when $y_1=y_2$, the integral over $y_1'$ becomes
 \begin{equation}\label{eq:linear-shear-Cvq-alpha-log-div}
\int \mbox{d}y'_{1}\,
\frac{H_{k}(y_{1},y'_{1})\mbox{e}^{-il(y'_{1}-y_{1})}}{\left(y_{1}-y'_{1}+\frac{2\alpha}{ks}i\right)^2}\,=
\int \mbox{d}y'_{1}\,
\frac{\frac{\partial }{\partial y_1'}\left(H_{k}(y_{1},y'_{1})\mbox{e}^{-il(y'_{1}-y_{1})}\right)}{y_{1}-y'_{1}+\frac{2\alpha}{ks}i}\,,
\end{equation}
where we used an integration by parts. We now see that this integral diverges when $\alpha\to0$ because the quantity at the numerator is not continuous exactly at $y_1=y_1'$. This implies that $D^\alpha_{kl}(y,y)$ diverges as $\ln\alpha$ when $\alpha\to0$\footnote{To understand the rate of this divergence with $\alpha$, it is enough to observe that the divergence arises from the neighbourhood $y_1'\in[y_1-\epsilon,y_1+\epsilon]$. Then, as $y_1'\to\frac{\partial }{\partial y_1'}\,H_{k}(y_{1},y'_{1})$ is analytic in both the neighbouroods $y_1'\in[y_1-\epsilon,y_1[$ and $y_1'\in]y_1,y_1+\epsilon]$, we can expand it in Taylor series. By direct computation, one finally obtains that the integral in \eqref{eq:linear-shear-Cvq-alpha-log-div} diverges as $\ln\alpha$.}.

Using \eqref{eq:linear-shear-Cvv-alpha} and the fact that $D^\alpha_{kl}(y,y)\sim\ln\alpha$, we deduce\footnote{See Appendix \ref{sec:SC-background-appendix-4pts} for details.} the asymptotic behaviour of the integrated autocorrelation function $\Xi^\alpha_{kl}(y_1,y_2)$ (defined in \eqref{eq:SC-Xi-Xi_kl})
\begin{equation}
\Xi^\alpha_{kl}(y_1,y_2) \;\underset{\alpha\to0}{\sim}\; \frac{A_{kl}(y_1,y_2)}{iks(y_{1}-y_{2})+2\alpha}
\end{equation}
where $A_{kl}(y_1,y_2)$ is a finite function independent of $\alpha$. We have thus proved the result \eqref{eq:SC-Xi-kl-convergence} in this simple case.



\section{Consequences for the dynamics of the jet velocity profile\label{sec:Consequences_mean_flow}}
We now discuss the consequences  of our analysis on the statistics of the large scale flow $U(y,t)$ and, where possible, compare them with the results obtained from direct numerical simulations. In paragraph \ref{sec:covariance-U}, we discuss the variance and the covariance of the jet velocity profile; in paragraph \ref{sec:SC-energy-Xi} we discuss the zonal energy balance, dividing the part contained in the average jet velocity profile and the one due to its fluctuations. 


\subsection{Covariance and variance of the jet velocity profile}\label{sec:covariance-U}
In order to analyse fluctuations of the jet velocity profile $U(y)$, we consider the covariance and variance of $U(y)$ in the stationary state, i.e., when the mean jet velocity profile is stationary. From a theoretical point of view, we consider the observables
\qq\label{eq:covariance-variance}
\mathbb{C}(y_1,y_2,t) = \mathbb{E}_K[U(y_1,t) \, U(y_2,t)] -U_d(y_1,t) \,U_d(y_2,t) \qquad \qquad 
\mathbb{V}(y,t) = \mathbb{C}(y,y,t)\,,
\qqq
where the average is taken over the noise appearing in the kinetic equation  (\ref{eq:kinetic_equation}) and $U_d(y,t)=\mathbb{E}_K[U(y,t)]$ is the mean jet velocity profile. In order to be precise, we have distinguished the average over the noise appearing in the kinetic equation (denoted by $\mathbb{E}_K$) from the average over the original noise appearing in the stochastic 2D Euler equations (denoted by $\mathbb{E}$).\\
The evolution equation for the covariance can be obtained straightforwardly from the kinetic equation (\ref{eq:kinetic_equation}) employing Ito's calculus and averaging over the noise:
\qq
\frac{d \mathbb{C}(y_1,y_2,t)}{dt}
 = -2\alpha \mathbb{C}(y_1,y_2,t)
+ \alpha^2 \mathbb{E}_K[\Xi[U](y_1,y_2,t)]
+\alpha \mathbb{E}_K[[ U(y_1,t) F^0[U](y_2,t) + F^0[U](y_1,t) U(y_2,t)  ]]\,
\qqq
and, thus, in a stationary state, 
\qq\label{eq:CV-int}
2\mathbb{C}(y_1,y_2,\infty)=
\lim_{t\to \infty}\left\{\alpha \mathbb{E}_K[\Xi[U](y_1,y_2,t)]
+ \mathbb{E}_K[[ U(y_1,t) F^0[U](y_2,t) + F^0[U](y_1,t) U(y_2,t)  ]]\right\}\,.
\qqq
As the kinetic equation for $U$ is non linear (in $U$), eq. \eqref{eq:CV-int} is not closed: the right hand side cannot be written as a function of the covariance because higher order correlations emerge. 

However, a striking qualitative feature emerges from a simple analysis. Assuming that $U-U_d$ small in some well suited norm, we can Taylor expand the right hand side of eq. (\ref{eq:CV-int}) up to second order. We then find that $\mathbb{C}$ is proportional to $\alpha \mathbb{E}_K[\Xi[U_d]]$. From the small $\alpha$ limit of the noise correlation, see eq. (\ref{eq:SC-Xi-kl-convergence}), we know that the first term of eq. (\ref{eq:CV-int}) is order $\alpha$ for all $y_1$ and $y_2$ except the following case: $U_d(y_1)=U_d(y_2)$ and $U'(y_1)\neq0$,  case in which it is of order one. Unless the non-linear terms involving the Reynold's stress exactly cancel out such behaviour, we  conclude that the same behaviour is expected for $\mathbb{C}(y_1,y_2,\infty)$. Moreover,  $\mathbb{V}(y,\infty)$ should be peaked when away from the stationary points of the average flow $U_d$.

The above argument is only qualitative. A more precise analysis would require, first of all, to check that the coefficients in the Taylor expansion are indeed finite. This would mean to analyse functional derivatives with respect to $U$ of the Reynolds stress $F^0$ and of  the noise covariance $\Xi$.  Beside the fact that such an analysis is likely to be technically cumbersome, it is not even clear that it would be useful. It is indeed hard to imagine how the qualitative picture obtained above might be changed, and it would not permit to get quantitative results anyway. The other possibility would be to perform numerical integration of the kinetic equation (\ref{eq:kinetic_equation}). Such second possibility is, however,  far from being straightforward and we consider it a very promising perspective of our work that we leave for future studies. 

Here, we content with the qualitative picture obtained above for $\mathbb{C}(y_1,y_2,\infty)$ and $\mathbb{V}(y,\infty)$ and compare them with results from direct numerical simulations. Such prediction is indeed striking and one might wonder weather it is confirmed by direct numerical simulations or, indeed, is an artifact of our theoretical approach.

\begin{figure}
\subfigure[\label{fig:UU_a1_lx07_beta0_N256}$|U_d(y_1)-U_d(y_2)|$, $\alpha=10^{-3}$]{\includegraphics[height=5cm]{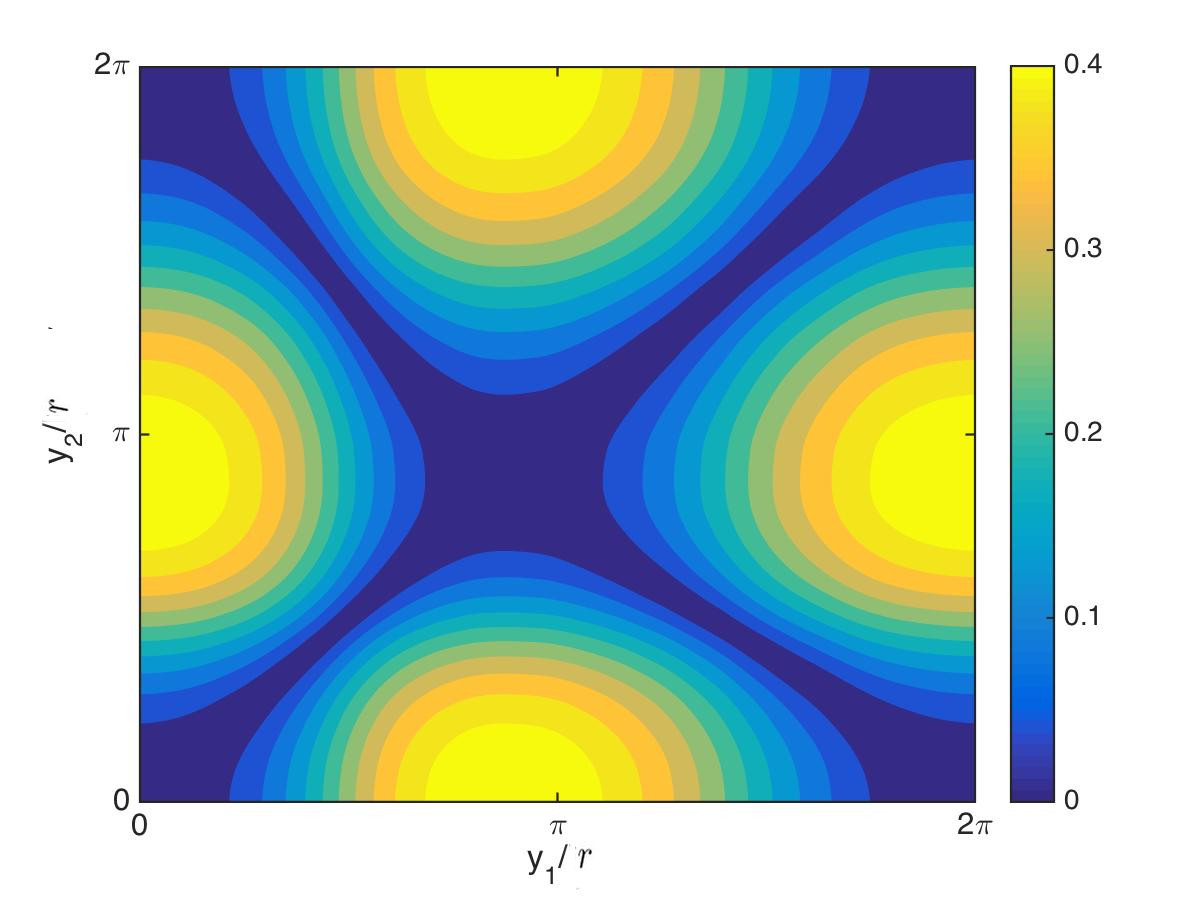}}
\subfigure[\label{fig:CV_a1_lx07_beta0_N256}$\mathrm{cov}(U(y_1),U(y_2))$, $\alpha=10^{-3}$]{\includegraphics[height=5cm]{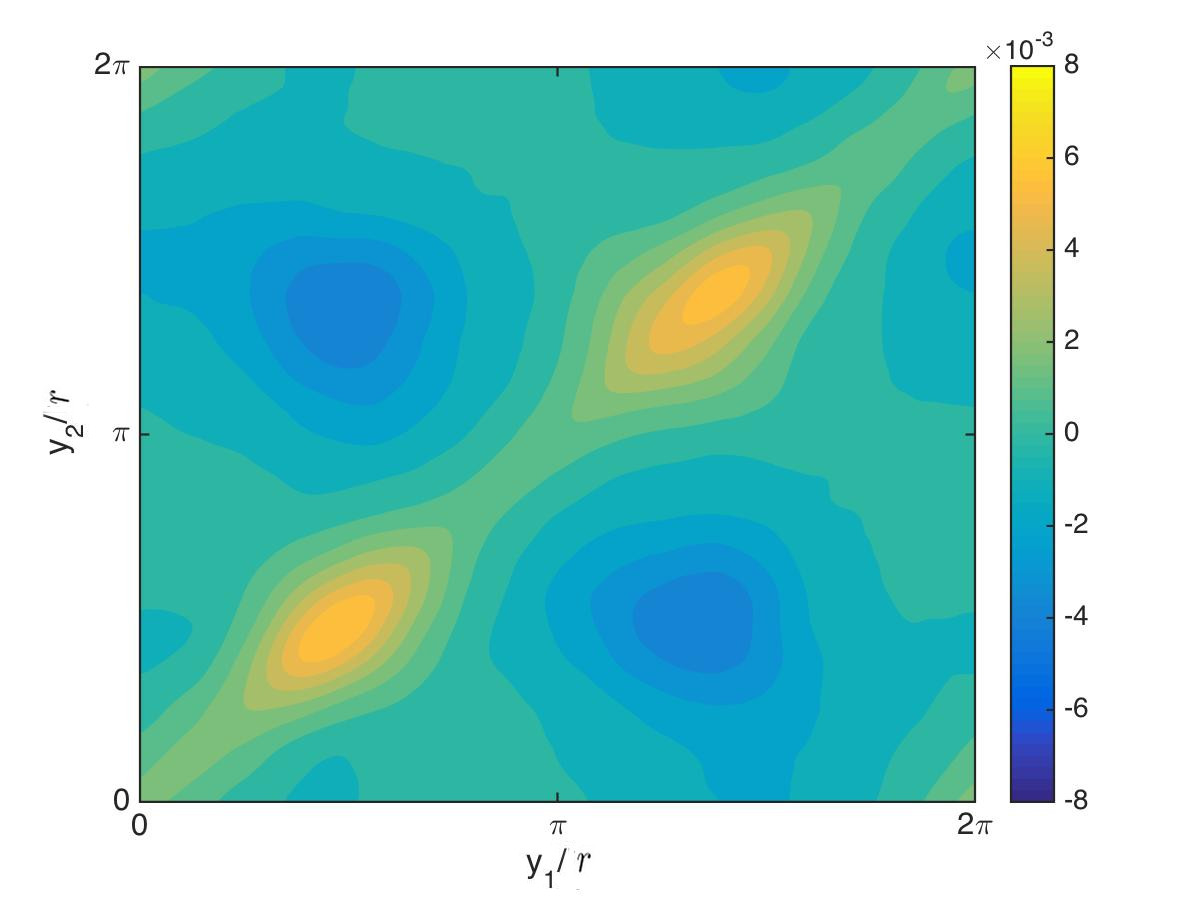}}\\
\subfigure[\label{fig:UU_a05_lx07_beta0_N256}$|U_d(y_1)-U_d(y_2)|$, $\alpha=5.10^{-4}$]{\includegraphics[height=5cm]{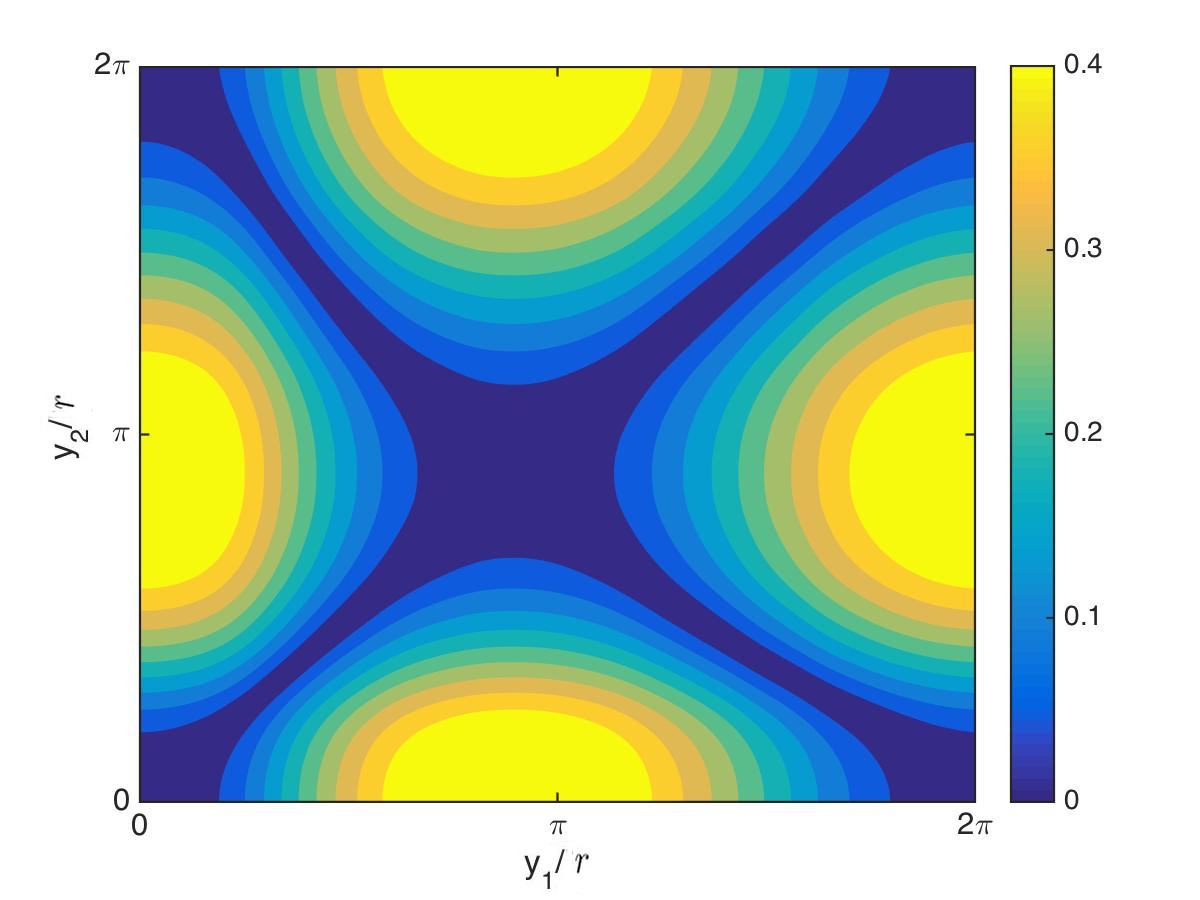}}
\subfigure[\label{fig:CV_a05_lx07_beta0_N256}$\mathrm{cov}(U(y_1),U(y_2))$, $\alpha=5.10^{-4}$]{\includegraphics[height=5cm]{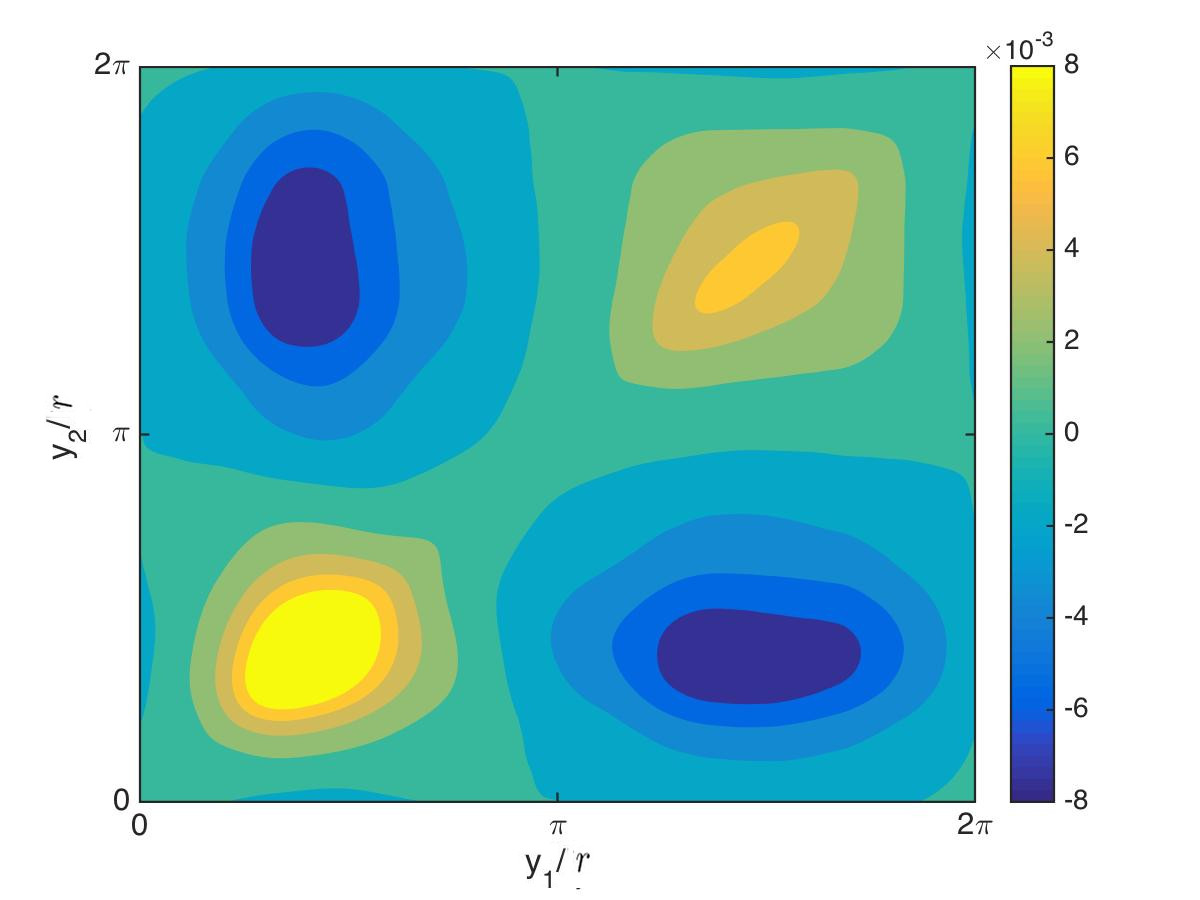}}\\
\caption{\label{fig:UU_vs_CV_lx07_beta0_N256}
(a) and (c): $|U_d(y_1)-U_d(y_2)|$ as a function of $(y_1,y_2)$ for (a) $\alpha=10^{-3}$ and (c) $\alpha=5.10^{-4}$, where $\bar{U}$ is the mean zonal velocity. (b) and (d): covariance of the zonal velocity $U$ as a function of $(y_1,y_2)$ for (b) $\alpha=10^{-3}$ and (d) $\alpha=5.10^{-4}$. Other parameters are the same as in figure \ref{fig:hovmoller_lx07_beta0_N256}. Large absolute values of the covariance are found in regions where $U(y_1)\simeq U(y_2)$ (along the diagonals). These regions also corresponds to maximal fluctuations of Reynolds stresses, according to the theoretical result (\ref{eq:SC-Xi-kl-convergence}). Along the diagonals, minimal absolute values of the covariance correspond to extrema of the mean zonal velocity, see figure \ref{fig:meanU_vs_varU_lx07_beta0_N256}.}\end{figure}

\begin{figure}
\subfigure[\label{fig:meanU_varU_a1_lx07_beta0_N256}$\alpha=10^{-3}$, $\nu_4=7.10^{-17}$]{\includegraphics[height=5cm]{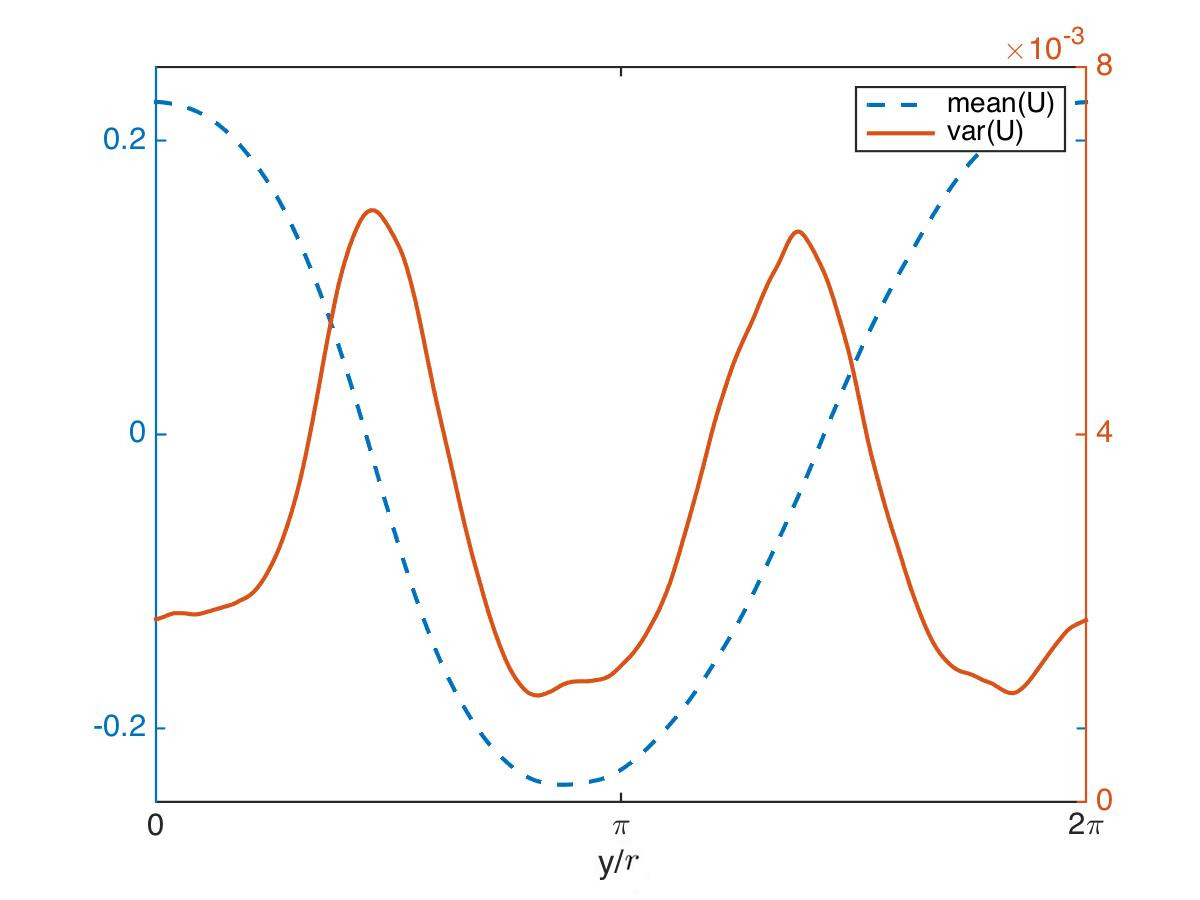}}
\subfigure[\label{fig:meanU_varU_a05_lx07_beta0_N256}$\alpha=5.10^{-4}$, $\nu_4=7.10^{-17}$]{\includegraphics[height=5cm]{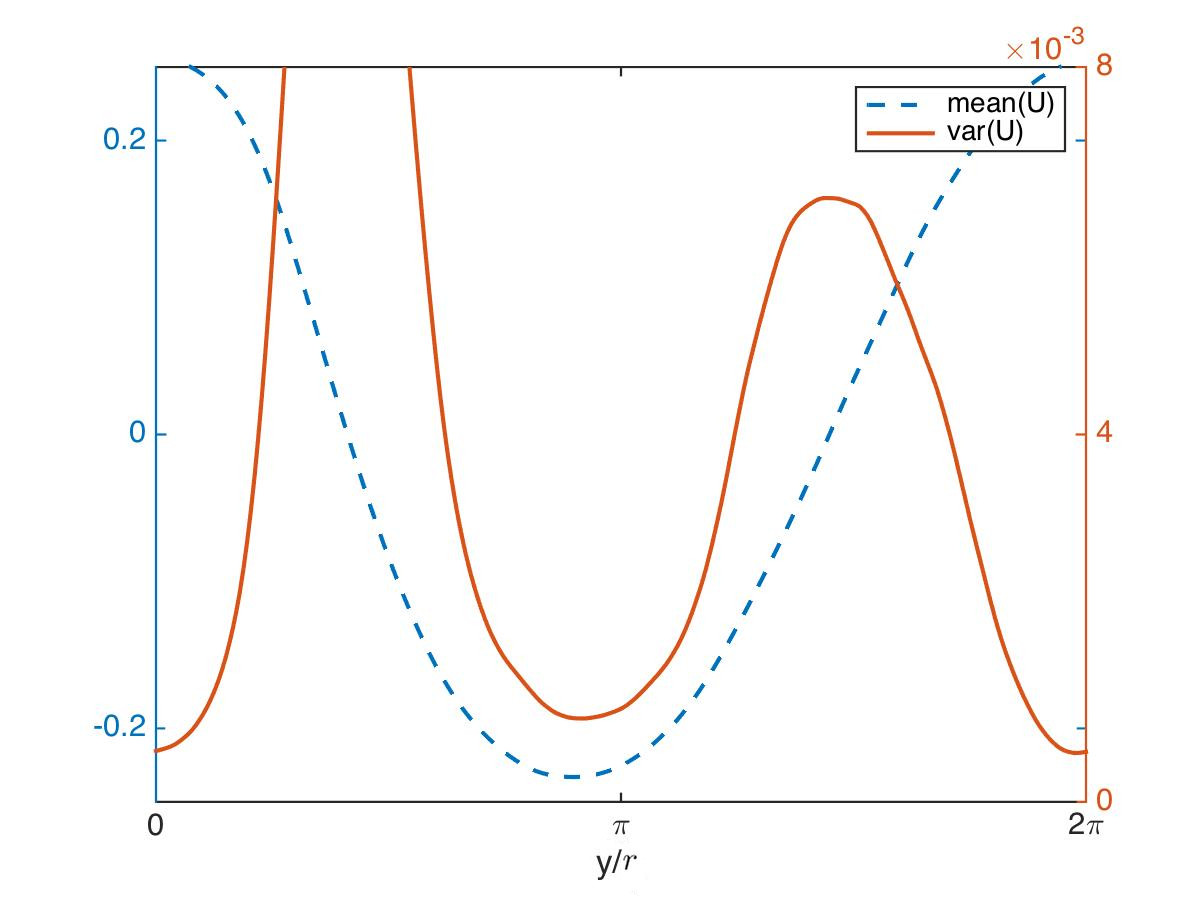}}\\
%
\caption{\label{fig:meanU_vs_varU_lx07_beta0_N256}Mean zonal velocity $\bar{U}(y)$ (dashed blue curve) and variance of the zonal velocity (solid orange curve) as functions of $y$ for (a) $\alpha=10^{-3}$ and (b) $\alpha=5.10^{-4}$. Other parameters are the same as in figure \ref{fig:hovmoller_lx07_beta0_N256}. For both values of $\alpha$, the regions of minimal variance correspond to extrema of the mean velocity $\bar{U}(y)$, i.e. $y/l_y \simeq 0$ and $y/l_y \simeq \pi$. This is a consequence of the depletion of vorticity at the stationary streamlines, which tends to reduce the fluctuations of Reynolds stresses in those regions compared to the fluctuations of Reynolds stresses in regions of shearing by the mean flow (here $y/l_y \simeq \pi/2$ and $y/l_y \simeq3\pi/2$). This numerical result is in qualitative agreement with the theoretical result (\ref{eq:SC-Xi-kl-convergence}).}\end{figure}

In Fig. \ref{fig:UU_vs_CV_lx07_beta0_N256}, the results for the stationary covariance $\mathbb{C}(y_1,y_2,\infty)$ of the jet velocity profile (on the right) is compared with the spatial structure of $|U_d(y_1)-U_d(y_2)|$. We reported results for the same resolution, forcing spectra, hyper-viscosity and two values of $\alpha$. Large absolute values of the covariance are found in regions where $U(y_1) \simeq U(y_2)$ (along the diagonals). These regions also corresponds to maximal fluctuations of Reynolds stresses, according to the theoretical result (3.41). Along the diagonals, minimal absolute values of the covariance correspond to extrema of the mean zonal velocity, see figure \ref{fig:meanU_vs_varU_lx07_beta0_N256}. We also report the results for smaller hyper-viscosity $\nu_4=3.5\times 10^{-17}$ and higher resolution ($512\times 512$), hence showing that our results do not depend on it. It is moreover clear from figure \ref{fig:meanU_vs_varU_lx07_beta0_N256} that diminishing $\alpha$ results in an increase of the variance everywhere except at the extrema of $U(y)$. These results are qualitatively consistent with the theoretical predictions we obtained from our kinetic theory. We are unfortunately unable to perform simulations on a sufficiently large range of $\alpha$ values in order to understand weather the $1/\alpha$ scaling of the $\mathbb{V}(y,\infty)$ from $y$ away from the stationary points of $U(y)$ is indeed present. 

\subsection{Energy balance for the large scales\label{sec:SC-energy-Xi}}
The kinetic energy 
associated with the jet is
\qq
\mathcal{E}_z[U] \equiv \pi l_x \int U^2(y)\,\mathrm{d} y\,,
\qqq
where the subscript $z$ stands for zonal and the average kinetic energy will be denoted by $E_z = \mathbb{E}_K[\mathcal{E}_z[U]]$. 
The evolution equation for $E_z$ is obtained 
applying the It\=o formula to the kinetic equation (\ref{eq:kinetic_equation})
\begin{equation}
\frac{dE_z}{dt} = 2\alpha \pi l_x \int \mathbb{E}_K\left[ F_0[U](y) U(y) \right] \,\mathrm{d} y -2\alpha E_z+ \alpha^2 \pi l_x \int \mathbb{E}_K\left[ \Xi[U](y,y) \right] \,\mathrm{d} y \,.
\label{eq:SC-zonal-energy-balance}
\end{equation}
Such equation is the average zonal energy balance: the first term represents the injection rate of energy in the large scales due to non-zonal degrees of freedom; the second term is Rayleigh friction; the last term, present only due to the noise in the kinetic equation \eqref{eq:kinetic_equation}, represents the energy injection rate in the zonal flow by the fluctuations of Reynolds stresses. Using our main theoretical result, eq. (\ref{eq:SC-Xi-kl-convergence}), we conclude that such term is actually of order $\mathcal{O}(\alpha)$ for $\alpha$ small. Our approach thus predict that the energy content in the fluctuations of $U$ is of the same order of magnitude of the energy contained in $U_d$.

It would be of interest to quantitatively test the above prediction by means of direct numerical simulations. This would require to study the scaling with $\alpha$ of the energy contained in the fluctuations of the jet velocity profile $\mathcal{E}_z[U] - \mathcal{E}_z[\mathbb{E}_K[U]]$. If the above prediction is correct, such quantity should remain of order $\mathcal{O}(1)$ in the small $\alpha $ limit. The problem is practically difficult to address because one has to be sure that hyper-viscosity is negligible and $\alpha$ small enough in order to be in the asymptotic regime; we were at the moment unable to get sufficiently clean results in order to conclude that the above statement is confirmed or disproved by direct numerical simulations.


\section{Conclusions\label{sec:Conclusion}}
Self-organisation in jets, i.e. flows that are mostly horizontal and unidirectional, is common in two-dimensional, quasi two-dimensional and geophysical turbulence. While much effort has been devoted in literature to the characterisation of their average velocity profile, little is instead known on the fluctuations, small and large, they undergo.

Some recent studies \citep{BS09,laurie2014langevin,bouchet2011control,rolland2016computing,wouters2015rare,laurie2015computation} concentrated on the description of large and abrupt fluctuations that large scale structures undergo. 
Examples of this behaviour are found in the magnetic field reversal for
the Earth, in MHD experiments \citep{monchaux2007}, in 3D flows \citep{ravelet2004}, in
atmospheric flows \citep{weeks1997}, oceaenic currents  \citep{schmeits2001}, and also in 2D turbulence
experiments \citep{sommeria1986}. The theoretical explanation of this behaviour is commonly done with large deviations techniques \citep{bouchet2015perturbative,bouchet2015large} but obtaining explicit theoretical results is a very difficult problem. Here, we are much more modest and concentrate on small, Gaussian, fluctuations close to the average state. \\

This paper is devoted to the study of Gaussian fluctuations of jet velocity profiles in the simplest possible theoretical framework: stochastic $2$-d Euler equations defined in eq. (\ref{eq:2D-NS-dimensional}). Our analysis is based on a non-equilibrium statistical mechanics approach first developed in \citep{Bouchet_Nardini_Tangarife_2013_Kinetic,nardini2014fluiddyn}, which has strong analogies with theories based on quasi-linear approximation such as CE2  \citep{Marston-2010-Chaos,Marston-APS-2011Phy,tobias2013direct,Marston_Conover_Schneider_JAS2008,Marston-APS-2011Phy,GormanSchneider-QL-GCM,Srinivasan-Young-2011-JAS,ait2015cumulant} and SSST \citep{Farrel_Ioannou,Farrell_Ioannou_JAS_2007,BakasIoannou2013SSST,ParkerKrommes2013SSST}. However, our approach goes beyond these theories, giving access not only to the average evolution of the jet velocity profile but also to the Gaussian fluctuations it undergoes. 

Once integrated out turbulent non-zonal fluctuations, the effective evolution for the jet velocity profile is expressed by eq. \eqref{eq:kinetic_equation}. Such effective evolution is expected to give very good predictions in the limit of negligible (hyper)-viscosity and when the the mean jet velocity profile evolves much slower than turbulent non-zonal fluctuations \citep{tobias2013direct,constantinou2014}. Such limit has been precisely discussed in section \ref{sec:slow_jet_dynamics} where, passing in non-dimensional units, it can be cast as $\nu_n\ll\alpha\ll1$. Here, $\nu_n$ is the properly non-dimensional (hyper)-viscosity and $\alpha$ is the ratio between the typical time scale for the advection of the small scales by the large scale jet and the typical time-scale for the evolution of the jet velocity profile.

It is remarkable that we could carry out most of our analysis analytically. Our central theoretical result is the characterisation of the spatial structure of the noise covariance \eqref{eq:Xi_U_alpha_def} entering in the effective evolution \eqref{eq:kinetic_equation}. Such characterisation is given by eq. \eqref{eq:SC-Xi-Xi_kl} and \eqref{eq:SC-Xi-kl-convergence}. It permitted to obtain rather surprising qualitative predictions on the fluctuations of the jet velocity profile $U(y)$. 

In particular, we analysed the covariance and the variance of the zonal velocity profile, defined in eq. (\ref{eq:covariance-variance}). If we were considering a system with a finite number of degrees of freedom, we would conclude that both the variance and the covariance are of order $\mathcal{O}(\alpha)$. However, we are dealing with a field problem with ultraviolet divergences. Then, we predicted that the stationary covariance scales as $\mathbb{C}(y_1,y_2,\infty) \sim \mathcal{O}(\alpha)$ except when $U_d(y_1)=U_d(y_2)$ and $U_d'(y_1)\neq 0$, where we have $\mathbb{C}(y_1,y_2,\infty) \sim \mathcal{O}(1)$. Moreover, the variance of the jet velocity profile is predicted to behave as $\mathbb{V}(y,\infty) \sim \mathcal{O}(1)$ unless $U_d'(y)=0$, in which case $\mathbb{V}(y,\infty) \sim \mathcal{O}(\alpha)$. For finite $\nu_n$ and $\alpha$, we thus predict where $\mathbb{V}(y,\infty) $ and $\mathbb{C}(y_1,y_2,\infty)$ should be enhanced with a precise spatial pattern. Mathematically, $\mathbb{C}(y_1,y_2,\infty) \sim \mathcal{O}(1)$ converges to a distribution for
 $\alpha\to0^+$.
 
\noindent Employing direct numerical simulations, we find a clear footprint of such a prediction, with the $\mathbb{C}(y_1,y_2,\infty)$ and $\mathbb{V}(y,\infty)$ presenting a spatial structure very similar to the one described above. Our results are summarised in figures \ref{fig:UU_vs_CV_lx07_beta0_N256} and \ref{fig:meanU_vs_varU_lx07_beta0_N256}. Unfortunately, a quantitative comparison of our theoretical results seems out of reach at the present day. In particular, 
we are unable to verify the correctness of the above scalings with $\alpha$. Moreover our theoretical results indicate that the energy contained in the fluctuations of the zonal jet velocity profile (see section \ref{sec:SC-energy-Xi}) should be of the same order of magnitude of the energy contained in the average velocity profile. This effect was neglected in \citep{Bouchet_Nardini_Tangarife_2013_Kinetic,nardini2014fluiddyn} but, again, we are unable to get a quantitative test of it by direct numerical simulations.
A careful computational check of these conclusions is very hard, as one has to work in the limit of negligible (hyper)-viscosity and very slow evolution. However, we believe that it might be clarified by means of efficient parallel codes and we leave it as a direction for future investigation. 

The present work is a first step toward the study of fluctuations of large scale structures two-dimensional and quasi two-dimensional turbulence. Many points remain indeed open. First of all, both our theoretical and numerical results are for the moment restricted to the stochastic Euler equations. This is motivated by the fact that the analytical work can be pushed very far in this case. It would be of certain interest to understand what is the effect of introducing a $\beta$-effect or bottom topography. Preliminary work in this direction suggest that our results remain unchanged at least when the $\beta$-effect or the topography is small. We will devote future work to the deepening of such an issue.


Secondly we have shown that, at order $\mathcal{O}(\alpha^2)$ in our perturbative expansion, ultraviolet divergences appear. It would be of great interest to understand whether the scaling with $\alpha$ of the stationary covariance of $U$ and of the energy contained in the fluctuations of $U$ precisely holds. One might indeed expect that ultraviolet divergences result in a non-trivial renormalization of such scaling. In this sense, a renormalization group approach \citep{chen1996renormalization} might point to a modification of the scaling with $\alpha$ of these quantities. The analysis of finite-dimensional systems, as those considered in \citep{hairer2009hot}, with homogenization techniques might shed some light.

Finally, a challenging perspective of our work is to numerically implement our effective evolution \eqref{eq:kinetic_equation}. This would give access to much more precise quantitative predictions for the evolution of the jet velocity profile. A very important avenue, in this sense, is to understand whether it is possible to develop numerical codes for the integration of the effective evolution \eqref{eq:kinetic_equation} that are significantly faster than the direct integration of the stochastic Euler equations.


\section*{Acknowledgements}
C. Nardini warmly acknowledges and remember T. Tangarife who, sadly, passed away few weeks after obtaining his Ph.D., when the present work was close to the end. For C. Nardini, Tom\'as was a precious collaborator and, more importantly, a friend. T. Tangarife and C. Nardini acknowledge F. Bouchet for several discussions during the development of this work as well for providing the first version of the pseudo-spectral code that has been used in the present paper. T. Tangarife and C. Nardini acknowledge A. Venaille for help about the practical implementation of direct numerical simulations.
This research has been supported by  (C. Nardini) ANR grant ANR STOSYMAP (ANR-2011-BS01-015) and by the EPSRC grant Nr. EP/J007404. T. Tangarife acknowledges funding from the European Research Council under European Union's Seventh Framework Programme (FP7/2007-2013 Grant Agreement no. 616811). Numerical results have been obtained with the PSMN platform in ENS-Lyon.



\appendix

\section{Proof of the convergence for any base flow\label{appendix:proof}}

In this appendix we study the behaviour for small $\alpha$ of the integrated autocorrelation function $\Xi^\alpha[U]$ defined in \eqref{eq:Xi_U_alpha_def}. More precisely, we prove \eqref{eq:SC-Xi-kl-convergence}. We recall the following definitions:
\begin{equation}
\Xi^\alpha[U](y_1,y_2)  =\sum_{(k,l)\in \mathbb{Z}^2} c_{kl}^2\, \big\lbrace\Xi^\alpha_{kl}(y_1,y_2)+\Xi^\alpha_{kl}(y_2,y_1)\big\rbrace
\label{eq:SC-Xi-Xi_kl-appendix}
\end{equation}
with $\Xi^\alpha_{kl}(y_1,y_2)=C^\alpha_{kl}(y_{1},y_{2},T_{0}=0)+D^\alpha_{kl}(y_{1},y_{2},T_{0}=0)$ where
\begin{equation}
C^\alpha_{kl}(y_{1},y_{2},T_{0})=\frac14\int_{T_0}^\infty \mathbb{E}_U^\alpha\left[  v_{kl}(y_1,s) v_{-k,-l}(y_2,0) \right]\mathbb{E}_U^\alpha\left[  \omega_{-k,-l} (y_1,s) \omega_{kl}(y_2,0) \right]\,\md s
\label{eq:C-vv-appendix}
\end{equation}
and
\begin{equation}
D_{kl}^\alpha(y_{1},y_{2},T_{0})=\frac14\int_{T_0}^\infty\mathbb{E}_U^\alpha\left[  v_{kl}(y_1,s) \omega_{-k,-l}(y_2,0) \right]\mathbb{E}_U^\alpha\left[   \omega_{-k,-l} (y_1,s) v_{kl}(y_2,0) \right]\,\md s.\label{eq:C-vq-appendix}
\end{equation}
For future simplicity, we have introduced the variable $T_{0}$. Indeed,
we are only interested in the large $s$ behavior of the integrands
above because there are no convergence problems around $s=0$. In the following, $T_{0}$ will
be fixed and assumed to be very large.\\

In sections \ref{appendix:Tomegaomega} and \ref{sec:SC-background-appendix-2pts} we study the large-$s$ behaviour of the two-points correlation functions $T^\alpha_{\omega\omega}$, $T^\alpha_{vv}$, $T^\alpha_{v\omega}$ and $T^\alpha_{\omega v}$ given by (\ref{eq:convergence-CR-qq}--\ref{eq:convergence-CR-qv}), using the Orr mechanism \eqref{eq:orr-mechanism-voricity-alpha}. Then, we will be able to study the small-$\alpha$ behaviour of $C^\alpha_{kl}$ and $D^\alpha_{kl}$ given by (\ref{eq:C-vv-appendix}, \ref{eq:C-vq-appendix}). This is done in section \ref{sec:SC-background-appendix-4pts}.

\subsection{Large time behavior of $T^\alpha_{\omega\omega}$\label{appendix:Tomegaomega}}

We report \eqref{eq:convergence-CR-qq} for convenience,
\begin{equation}
T^\alpha_{\omega\omega}(k,l,y_{1},y_{2},s)\equiv
\frac12\mathbb{E}_{U}^\alpha\left[\omega_{-k,-l}(y_{1},s)\omega_{k,l}(y_{2},0)\right]=
\int_{0}^{\infty}\mbox{d}t_1\,\tilde{\omega}_{-k,-l}(y_{1},s+t_1)\tilde{\omega}_{k,l}(y_{2},t_1)\,,
\end{equation}
where $\tilde{\omega}_{k,l}(y,t)$ is the solution of the deterministic dynamics $\partial_t+L_{U,k}^0+\alpha$ with initial condition $\tilde{\omega}_{k,l}(y,0) = \me^{ily}$. Alternatively, we can write
\begin{equation}
T^\alpha_{\omega\omega}(k,l,y_{1},y_{2},s)=
\int_{0}^{\infty}\mbox{d}t_1\,\tilde{\omega}^*(y_{1},s+t_1)\tilde{\omega}(y_{2},t_1)\me^{-2\alpha t_1}\,,
\label{eq:def_T_omega_omega_alpha_appendix}
\end{equation}
where $\tilde{\omega}(y,t)$ is the solution of the deterministic inertial dynamics $\partial_t+L_{U,k}^0$ with initial condition $\tilde{\omega}_{k,l}(y,0) = \me^{ily}$, for simplicity in the notations we stop denoting the $(k,l)$ dependency and we also denote $T^\alpha(y_{1},y_{2},s)\equiv T^\alpha_{\omega\omega}(k,l,y_{1},y_{2},s)$.
We will prove that
\begin{equation}
T^\alpha(y_{1},y_{2},s) =  \frac{\tilde{\omega}_{-k,-l}^{\infty}(y_{1})\tilde{\omega}_{kl}^{\infty}(y_{2})}{ik\left(U(y_{1})-U(y_{2})\right)+2\alpha}\mbox{e}^{ikU(y_ {1})s-\alpha s}+T^{r,\alpha}(y_{2},y_{2},s)\,,
\label{eq:SC-Tqq-computation-appendix}
\end{equation}
where $T^{r,\alpha}(y_{2},y_{2},s)$ is finite for all $(y_{2},y_{2})$ such that $U(y_1)\neq U(y_2)$, and is negligible with respect to $1/\alpha$ if $U(y_1)= U(y_2)$. Also, $T^{r,\alpha}(y_{2},y_{2},s)$ is a bounded function of $s$.

\subsubsection{Resolvant of the linearized Euler operator\label{sec:SC-resolvant-euler-appendix}}

In order to prove \eqref{eq:SC-Tqq-computation-appendix}, we need to give a more complete version of the Orr mechanism. This is briefly presented in this paragraph, which is a reproduction of the technical results of \citep{Bouchet_Morita_2010PhyD}.\\

We define the Laplace transform of the deterministic vorticity as
\begin{equation}
\hat{\omega}(y,c+i\epsilon)=\int_{0}^{\infty}\mbox{d}t\,\tilde{\omega}(y,t)\mbox{e}^{ik(c+i\epsilon)t}.
\label{Laplace}
\end{equation}
The inverse Laplace transform is then given by
\begin{equation}
\tilde{\omega}(y,t)=\frac{|k|}{2\pi}\lim_{\epsilon\to0^+}\int_{-\infty}^{\infty}\mbox{d}c\,\hat{\omega}(y,c+i\epsilon)\mbox{e}^{-ikct}.
\label{eq:deterministic-vorticity-alpha}
\end{equation}
It is also useful to define the Laplace transform of the stream function $\phi\equiv\Delta^{-1}\hat{\omega}$; this quantity is usually referred in literature as the resolvent of the operator $L_{U,k}$. It is related to $\hat{\omega}$
through $\hat{\omega}(y,c+i\epsilon)=\left(\frac{d^{2}}{dy^{2}}-k^{2}\right)\phi(y,c+i\epsilon)$
and is the solution of the linear ordinary differential equation
\begin{equation}
\left(\frac{d^{2}}{dy^{2}}-k^{2}\right)\phi-\frac{U''(y)}{U(y)-c-i\epsilon}\phi=\frac{\mbox{e}^{ily}}{ik\left(U(y)-c-i\epsilon\right)}.
\label{eq:Rayleigh-equation}
\end{equation}
The homogeneous part of this equation (with zero right-hand side) is known as the Rayleigh equation \citep{drazinreid1981}.
For all $\epsilon>0$, this equation is a regular ODE. When $\epsilon\to0^{+}$,
this equation becomes singular at the critical layer $c=U(y)$. It
can be shown that $\phi(y,c+i\epsilon)\to\phi_{+}(y,c)$ as $\epsilon\to0^{+}$,
where $\phi_{+}$ is continuous over $c\in\mathbb{R}$, with either a logarithmic singularity in its first derivative with respect to $c$ if $U'(y)\neq0$, or a logarithmic singularity in its second derivative if $U'(y)=0$ \citep{Bouchet_Morita_2010PhyD}. We will first consider the case $U'(y)\neq0$. Then we can write, for all $c$,
\begin{equation}
\phi_{+}(y,c)=\phi_{2}(y,c).(U(y)-c)\ln\left|U(y)-c\right|+\phi_{1}(y,c),\label{eq:resolvant-all-c}
\end{equation}
where $\phi_1,\phi_{2}$ are analytic functions of $c$ \citep{Bouchet_Morita_2010PhyD}.\\

Using \eqref{eq:deterministic-vorticity-alpha}, \eqref{eq:Rayleigh-equation} and Plemelj formula \eqref{eq:Plemelj} to evaluate the limit $\epsilon\to0^+$, we get
\begin{equation}
\tilde{\omega}(y,t_{1})=\tilde{\omega}^{\infty}(y)\mbox{e}^{-ikU(y)t_{1}}+\int^{*}\frac{\mbox{d}c}{2\pi i}\,\frac{ikU''(y)\phi_+(y,c)+\mbox{e}^{ily}}{U(y)-c}\mbox{e}^{-ikct},\label{eq:deterministic-vorticity-Plemelj}
\end{equation}
where
\begin{equation}
\tilde{\omega}^{\infty}(y)=ikU''(y)\phi_{+}(y,U(y))+\mbox{e}^{ily},
\label{eq:SC-asymptotic-vorticity-appendix}
\end{equation}
and where we recall that $\int^*$ denotes the Cauchy Principal Value of the integral.

The first term is the classical Orr mechanism \eqref{eq:orr-mechanism-voricity}, the second term decays
for large $t_{1}$ as $1/t_{1}^{\gamma}$, where $\gamma>0$ depends
on the order of differentiability of $c\to\phi_{+}(y,c)$. Then, as $c\to\phi_+(y,c)$ is smoother at points $y$ such that $U'(y)=0$ than at points such that $U'(y)\neq0$, we can focus on the latter case.

\subsubsection{Points such that $U(y_1)=U(y_2)$}

The Orr mechanism \eqref{eq:deterministic-vorticity-Plemelj} can be written $\tilde\omega(y,t_1) = \tilde\omega^\infty(y) \mbox{e}^{-ikU(y)t_{1}} + \tilde\omega^r(y,t_1)$ where $\tilde\omega^r(y,t_1)= O(t_1^{-\gamma})$ as $t_1\to\infty$, with $\gamma>0$. Using this expression of the deterministic vorticity and \eqref{eq:def_T_omega_omega_alpha_appendix}, we get
\[T^\alpha(y_{1},y_{2},s)= \frac{\tilde\omega^\infty(y_1)\tilde\omega^{\infty*}(y_2)}{ik(U(y_1)-U(y_2))+2\alpha}+ \tilde{g}^{r}(y_1,y_2)\,,\]
where $\tilde{g}^{r}$ contains the corrections involving $\tilde\omega^r$. From Plemelj formula \eqref{eq:Plemelj}, the first term converges to a distribution in the limit $\alpha\to0^+$. In particular, it diverges point-wise as $1/\alpha$ at points such that $U(y_1)=U(y_2)$. We now prove that the remainder $\tilde{g}^{r}$ is negligible compared to this $1/\alpha$ divergence at such points.

The most divergent part of $\tilde{g}^{r}$ is of the form $G_\alpha = \int_0^\infty f(t)\,\md t $ where $f$ is bounded and $f(t)= O(\me^{-2\alpha t}t^{-\gamma})$ as $t\to\infty$. The behaviour for small $\alpha$ of $G_\alpha$ depends on the value of $\gamma$.
\begin{itemize}
\item if $\gamma<1$, there exists some $K>0$ such that
\[|G_\alpha |\leq K\int_0^\infty \frac{\me^{-2\alpha t}}{t^\gamma}\,\md t,\]
which is finite for all $\alpha>0$ because the integrand is integrable close to $t=0$. With the change of variable $u=\alpha t$, we get $|G_\alpha| \leq K' \alpha^{\gamma-1}$ with $K'=K\int_0^\infty \me^{-2 u}u^{-\gamma}\,\md u$.
\item if $\gamma=1$, taking the derivative with respect to $\alpha$ and with the change of variable $u=\alpha t$ we get
\[\frac{\partial G_\alpha}{\partial\alpha} = \frac{1}{\alpha}\int_0^\infty \me^{-2u}g\left(\frac{u}{\alpha}\right)\,\md u\]
with a bounded function $g$ such that $g(t)=O(1)$ as $t\to\infty$. By the theorem of dominated convergence,
\[\int_0^\infty \me^{-2u}g\left(\frac{u}{\alpha}\right)\,\md u \;\underset{\alpha\to0}{\longrightarrow}\; K''\equiv \int_0^\infty \me^{-2u}\lim_\infty g\,\md u \,,\]
so by integration $G_\alpha\sim-K''\ln\alpha$.
\item if $\gamma>1$ we directly have $G_\alpha\to G_0<\infty$ as $\alpha\to0$ by the theorem of dominated convergence.
\end{itemize}
In all three cases, $G_\alpha$ is negligible with respect to $1/\alpha$ as $\alpha\to0$. We conclude that $\tilde{g}^r(y_1,y_2)$ is negligible with respect to the $1/\alpha$ divergence of $\tilde{g}(y_1,y_2)$ at points such that $U(y_1)=U(y_2)$.

At points such that $U(y_1)\neq U(y_2)$, the first term in the expression of $\tilde{g}(y_1,y_2)$ converges to a finite value, so we need to prove that $\tilde{g}^r(y_1,y_2)$ also converges. This is done in next paragraph.

\subsubsection{Points such that $U(y_1)\neq U(y_2)$}
Let us now consider 
\begin{equation}
\tilde{g}(y_{1},y_{2})=\int_0^\infty \tilde{\omega}(y_1,t_1)\tilde{\omega}^*(y_2,t_1)\me^{-2\alpha t_1}\,\md t_1.
\label{eq:gkl-omega-tilde}
\end{equation}
Using \eqref{eq:deterministic-vorticity-alpha} and \eqref{eq:Rayleigh-equation}
we get
\begin{equation}
\tilde{g}(y_{1},y_{2})=\lim_{\epsilon_{1},\epsilon_{2}\to0^{+}}\int\frac{\mbox{d}c_{1}}{2\pi}\frac{\mbox{d}c_{2}}{2\pi}\,\hat{\omega}(y_{1},c_{1}+i\epsilon_{1})\left(\hat{\omega}(y_{2},c_{2}+i\epsilon_{2})\right)^{*}\frac{1}{ik(c_{1}-c_{2})+2\alpha}\label{eq:gkl-integral-c1-c2}
\end{equation}
with $\hat{\omega}(y,c+i\epsilon)=\frac{ikU''(y)\phi(y,c+i\epsilon)+\mbox{e}^{ily}}{U(y)-c-i\epsilon}$.
We easily realize that the infinite bounds
of this double integral are not sources of divergence. The only possible
sources of divergence come from the critical layers $c=U(y)$ when
$\alpha\to0^{+}$. When $U(y_{1})=U(y_{2})$, we know that $\tilde g$
is equivalent to $1/\alpha$ as $\alpha\to0$. We now consider the case $U(y_{1})\neq U(y_{2})$.

Consider a fixed $\alpha>0$, $\tilde g$ is of the form 
\[
I_{\alpha}=\lim_{\epsilon_{1}\to0^{+}}\int\mbox{d}x_{1}\,\frac{f_{1}(x_{1})}{x_{1}-a_{1}-i\epsilon_{1}}\lim_{\epsilon_{2}\to0^{+}}\int\mbox{d}x_{2}\,\frac{f_{2}(x_{2})}{x_{2}-a_{2}-i\epsilon_{2}}\frac{1}{x_{1}-x_{2}-i\alpha},
\]
where the functions $x\to f_{k}(x)$ are continuous with a logarithmic singularity
in their first derivative at $x=a_{k}$. We also assume that $a_{1}\neq a_{2}$.
Using Plemelj formula \eqref{eq:Plemelj} to estimate successively the limits $\epsilon_{2}\to0^{+}$
and $\epsilon_{1}\to0^{+}$, we get
\begin{equation}
\begin{aligned}
I_{\alpha} & =\frac{\pi^{2}f_{1}(a_{1})f_{2}(a_{2})}{a_{1}-a_{2}-i\alpha}-i\pi f_{2}(a_{2})\int^{*}\mbox{d}x_{1}\,\frac{f_{1}(x_{1})}{x_{1}-a_{1}}\frac{1}{x_{1}-a_{2}-i\alpha}\\
 & -i\pi f_{1}(a_{1})\int^{*}\mbox{d}x_{2}\,\frac{f_{2}(x_{2})}{x_{2}-a_{2}}\frac{1}{a_{1}-x_{2}-i\alpha}\\
 & -\int^{*}\mbox{d}x_{1}\,\frac{f_{1}(x_{1})}{x_{1}-a_{1}}\int^{*}\mbox{d}x_{2}\,\frac{f_{2}(x_{2})}{x_{2}-a_{2}}\frac{1}{x_{1}-x_{2}-i\alpha},\\
\end{aligned}\label{eq:SC-I-alpha-appendix}\end{equation}
where all the principal value integrals are finite because $f_{1}$
and $f_{2}$ are continuous, and because $\alpha>0$. We now study
the convergence of each term as $\alpha\to0^{+}$.
\begin{itemize}
\item The first term $\frac{\pi^{2}f_{1}(a_{1})f_{2}(a_{2})}{a_{1}-a_{2}-i\alpha}$
converges to $\frac{\pi^{2}f_{1}(a_{1})f_{2}(a_{2})}{a_{1}-a_{2}}$,
which is finite for $a_{1}\neq a_{2}$. This term corresponds to the most divergent part when $a_1=a_2$ (or $U(y_1)=U(y_2)$ in $\tilde g$). It also means that the convergence of the remaining terms in \eqref{eq:SC-I-alpha-appendix} depends directly on the value of $\gamma$ in the Orr mechanism, or equivalently on the regularity of the resolvant $c\to \phi_+(y,c)$. 
\item For the second term, Plemelj
formula \eqref{eq:Plemelj} can be applied to estimate the limit $\alpha\to0^{+}$ because
the singularities at $x_{1}=a_{1}$ and $x_{1}=a_{2}$ are not confounded:
\[\int^{*}\mbox{d}x_{1}\,\frac{f_{1}(x_{1})}{x_{1}-a_{1}}\frac{1}{x_{1}-a_{2}-i\alpha} \;\underset{\alpha\to0^+}{\longrightarrow}\; \pi\frac{f_{1}(a_{2})}{a_{2}-a_{1}} - i \int^{*}\mbox{d}x_{1}\,\frac{f_{1}(x_{1})}{(x_{1}-a_{1})(x_{1}-a_{2})}\]
The same result applies to the third term.
\item For the last term, let's
consider the function
\[
J(x_{1})=\lim_{\alpha\to0^{+}}\int^{*}\mbox{d}x_{2}\,\frac{f_{2}(x_{2})}{x_{2}-a_{2}}\frac{1}{x_{1}-x_{2}-i\alpha}.
\]
At any point such that $x_{1}\neq a_{2}$, this can be estimated using
Plemelj formula \eqref{eq:Plemelj},
\[
J(x_{1})=\frac{\pi f_{2}(x_{1})}{x_{1}-a_{2}}-i\int^{*}\mbox{d}x_{2}\,\frac{f_{2}(x_{2})}{(x_{2}-a_{2})(x_{1}-x_{2})},
\]
where both terms are finite because $x_{1}\neq a_{2}$. To estimate
the limit at the point $x_{1}=a_{2}$, we first use that
\begin{equation}
\int^* \frac{f(y)}{y}\,\md y  = \int \frac{f(y)-f(0)}{y}\,\md y \,,
\label{eq:SC-Cauchy-Riemann-appendix}
\end{equation}
where the integral on the right-hand side is now a usual Riemann integral if $f$ is continuous at $y=0$. This equality indeed follows from
\[ \int^* \frac{f(y)}{y}\,\md y \equiv \int PV\left(\frac{1}{y}\right)f(y) \,\md y \equiv \lim_{\epsilon\to0^+} \left[\int_{-\infty}^{-\epsilon} + \int_\epsilon^{+\infty} \right] \frac{f(y)}{y}\,\md y \,\]
and the fact that $\int^*\md y/y = 0$. We thus have 
\[
J(a_{2})=\lim_{\alpha\to0^{+}}\int\mbox{d}x_{2}\,\frac{1}{x_{2}-a_{2}}\left(\frac{f_{2}(x_{2})}{a_{2}-x_{2}-i\alpha}-\frac{f_{2}(a_{2})}{-i\alpha}\right)
\]
and the expression of the resolvant \eqref{eq:resolvant-all-c}, here $f_{2}(x_{2})=g(x_{2})(x_{2}-a_{2})\ln\left|x_{2}-a_{2}\right|+h(x_{2})$,
where $g$ and $h$ are analytic functions,
\begin{equation}
J(a_{2})=-\lim_{\alpha\to0^{+}}\left[\int\mbox{d}x_{2}\,\frac{g(x_{2})\ln\left|x_{2}-a_{2}\right|}{x_{2}-a_{2}+i\alpha}+\int^{*}\mbox{d}x_{2}\,\frac{h(x_{2})}{x_{2}-a_{2}}\frac{1}{x_{2}-a_{2}+i\alpha}\right],
\label{eq:cesare-1}
\end{equation}
where we have used again \eqref{eq:SC-Cauchy-Riemann-appendix} in order to express the second integral as a Principal Value. 
The first term in the brackets finite for all $\alpha>0$ because $x\to\ln x$ is
integrable around $x=0$. This
term converges in the limit $\alpha\to0$:
\begin{align*}
\int\mbox{d}x_{2}\,\frac{g(x_{2})\ln\left|x_{2}-a_{2}\right|}{x_{2}-a_{2}+i\alpha} & \underset{\alpha\to0^{+}}{\sim}\int\mbox{d}x_{2}\,\frac{g(x_{2})\ln\left|x_{2}-a_{2}+i\alpha\right|}{x_{2}-a_{2}+i\alpha}\\
 & =\int\mbox{d}x_{2}\, g(x_{2})\frac{1}{2}\frac{d}{dx_{2}}\ln^{2}\left|x_{2}-a_{2}+i\alpha\right|\\
 & =-\frac{1}{2}\int\mbox{d}x_{2}\, g'(x_{2})\ln^{2}\left|x_{2}-a_{2}+i\alpha\right|\\
 & \underset{\alpha\to0^{+}}{\to}-\frac{1}{2}\int\mbox{d}x_{2}\, g'(x_{2})\ln^{2}\left|x_{2}-a_{2}\right|,
\end{align*}
where the first and last equivalents follow from continuity of $z\to \ln|z|$. This expression is finite because $g$ is analytic and $x\to\ln^{2}(x)$ is
integrable around $x=0$. In the second term in \eqref{eq:cesare-1}, we use that $h$ can be expanded in its Taylor series,
$h(x_{2})=h_{0}+h_{1}(x_{2}-a_{2})+o(x_{2}-a_{2})$, so
\[
\int^{*}\mbox{d}x_{2}\,\frac{h(x_{2})}{x_{2}-a_{2}}\frac{1}{x_{2}-a_{2}+i\alpha}=\int^{*}\mbox{d}x_{2}\,\frac{h_{0}}{x_{2}-a_{2}}\frac{1}{x_{2}-a_{2}+i\alpha}+\int\mbox{d}x_{2}\,\frac{h_{1}+o(1)}{x_{2}-a_{2}+i\alpha},
\]
where the last integral is now a usual Riemann integral because the divergence has been cancelled. The term involving $h_{0}$ can be computed
explicitely for any $\alpha>0$, 
\[
\int^{*}\mbox{d}x_{2}\,\frac{1}{x_{2}-a_{2}}\frac{1}{x_{2}-a_{2}+i\alpha}=\frac{1}{i\alpha}\lim_{\epsilon\to0^{+}}\left[\ln\left|x_{2}-a_{2}+i\alpha\right|-\ln\left|x_{2}-a_{2}\right|\right]_{a_{2}+\epsilon}^{a_{2}-\epsilon}=0.
\]
Then,
\[
\int^{*}\mbox{d}x_{2}\,\frac{h(x_{2})}{x_{2}-a_{2}}\frac{1}{x_{2}-a_{2}+i\alpha} = \int\mbox{d}x_{2}\,\frac{h(x_2)-h(a_2)}{x_{2}-a_{2}+i\alpha}\;\underset{\alpha\to0^+}{\longrightarrow}\;  \int\mbox{d}x_{2}\,\frac{h(x_2)-h(a_2)}{x_{2}-a_{2}},
\]
which is finite. We conclude that $J(x_{1})$ is a finite quantity for all $x_{1}$, and is continuous at $x_1=a_1\neq a_2$.
Then,
\[
\lim_{\alpha\to0^{+}}\int^{*}\mbox{d}x_{1}\,\frac{f_{1}(x_{1})}{x_{1}-a_{1}}\int^{*}\mbox{d}x_{2}\,\frac{f_{2}(x_{2})}{x_{2}-a_{2}}\frac{1}{x_{1}-x_{2}-i\alpha}=\int^{*}\mbox{d}x_{1}\,\frac{f_{1}(x_{1})}{x_{1}-a_{1}}J(x_{1}),
\]
which is finite.
\end{itemize}
We conclude that $I_{\alpha}$ has a finite limit
for $\alpha\to0^{+}$, so $\tilde{g}(y_{1},y_{2})$ is finite for all
points such that $U(y_{1})\neq U(y_{2})$.

\subsection{Other two--points correlation functions\label{sec:SC-background-appendix-2pts}}

\subsubsection{Large time behavior of $T^\alpha_{vv}$}

We report \eqref{eq:convergence-CR-vv} for convenience,
\begin{equation}
T^\alpha_{vv}(k,ly_{1},y_{2},s)\equiv
\frac12\mathbb{E}_{U}^\alpha\left[v_{k,l}(y_{1},s)v_{-k,-l}(y_{2},0)\right]=
 \int_{0}^{\infty}\mbox{d}t_1\,\tilde{v}_{k,l}(y_{1},s+t_1)\tilde{v}_{-k,-l}(y_{2},t_1)\,.
\end{equation}
We show here that $T^\alpha_{vv}$ decays as or faster than $1/s^2$.\\
We have
\begin{equation}
\left|T^\alpha_{vv}(k,l,y_{1},y_{2},s)\right|\leq \int_{0}^{\infty}\mbox{d}t_1\,\left|\tilde{v}_{k,l}(y_{1},t_1+s)\right|\left|\tilde{v}_{-k,-l}(y_{2},t_1)\right|\,.
\end{equation}
Because $T_{0}\gg1$, we can chose in the above formula $s\gg1$.
We thus have
\begin{equation}
\left|T^\alpha_{vv}(k,l,y_{1},y_{2},s)\right|\leq\left|\frac{\tilde{\omega}_{k,l}^{\infty}(y)}{ik(U'(y))^{2}}\right|\int_{0}^{\infty}\mbox{d}t_1\,\left|\tilde{v}_{-k,-l}(y_{2},t_1)\right|\left\{ \frac{1}{(t_1+s)^{2}}+o\left(\frac{1}{(t_1+s)^{2}}\right)\right\} \,.
\end{equation}
Using the results in section \ref{sec:appendix-Temporal-decay-of},
we have
\begin{equation}
\left|T^\alpha_{vv}(k,l,y_{1},y_{2},s)\right|\underset{t\to\infty}{\lesssim}\frac{R_{vv}(k,l,y_{1},y_{2})}{s^{2}}\,+o\left(\frac{1}{s^{2}}\right),
\label{eq:Tvv-asymptotic-behaviour}\end{equation}
where $R_{vv}$ is a positive, bounded function of $(y_{1},y_{2})$.
It is important to note that $R_{vv}$ does not depend on $\alpha$.

\subsubsection{Large time behavior of $T^\alpha_{v\omega}$}
We report \eqref{eq:convergence-CR-vq} for convenience,
\begin{equation}
T^\alpha_{v\omega}(k,l,y_{1},y_{2},s)\equiv
\frac12\mathbb{E}_{U}^\alpha\left[v_{k,l}(y_{1},s)\omega_{-k,-l}(y_{2},0)\right]=
\int_{0}^{\infty}\mbox{d}t_1\,\tilde{v}_{k,l}(y_{1},s+t_1)\tilde{\omega}_{-k,-l}(y_{2},t_1)\, .
\end{equation}
The large-$s$ behavior of $T^\alpha_{v\omega}(k,l,y_1,y_2,s)$ is different if $U(y_{1})=U(y_{2})$
or if $U(y_{1})\neq U(y_{2})$. Indeed, in the first case, the asymptotic
oscillations of the integral cancel out and the large-$s$ decay is
slower: it decays as $1/s$ in the $\alpha\to0$ limit. In the second
one, the oscillations do not cancel out and the decay is as $1/s^{\min\{1+\gamma,2\}}$, where $\gamma>0$ is the exponent of the
decay of $\tilde{\omega}^{r}_{kl}$.

We have
\begin{equation}\begin{aligned}
T^\alpha_{v\omega}(k,l,y_{1},y_{2},s)\underset{t\to\infty}{\sim}\frac{\tilde{\omega}_{k,l}^{\infty}(y_{1})}{ik(U'(y_{1}))^{2}}\mbox{e}^{-ikU(y_{1})s-\alpha s}&\left\{ \tilde{\omega}_{-k,-l}^{\infty}(y_{2})\int_{0}^{\infty}\mbox{d}t_1\,\frac{\me^{-i\left[kU(y_{1})-kU(y_{2})\right]t_1-2\alpha t_1}}{(t_1+s)^{2}}\right.\\
&\left.+\int_{0}^{\infty}\mbox{d}t_1\,\frac{\mbox{e}^{-ikU(y_{1})t_1-2\alpha t_1}}{(t_1+s)^{2}}\tilde{\omega}_{-k,-l}^{r}(y_{2},t_1)\right\} \,.\\
\end{aligned}
\label{eq:3.2.3-correlation-vq-explicit-2}
\end{equation}
We now see that the decay in $s$ of the expression in parenthesis
is different if $U(y_{1})=U(y_{2})$ or $U(y_{1})\neq U(y_{2})$.

If $U(y_{1})=U(y_{2})$, the first integral dominates. We have 
\begin{equation}
\int_{0}^{\infty}\mbox{d}t_1\,\frac{\me^{-i\left[kU(y_{1})-kU(y_{2})\right]t_1-2\alpha t_1}}{(t_1+s)^{2}}=\int_{0}^{\infty}\mbox{d}t_1\,\frac{\me^{-2\alpha t_1}}{(t_1+s)^{2}}\leq\frac{1}{s}.
\end{equation}
Observe that the equality holds in the $\alpha\to0$ limit. We conclude
that, if $U(y_{1})=U(y_{2})$, 
\begin{equation}
\left|T^\alpha_{v\omega}(k,l,y_{1},y_{2},s)\right|\underset{s\to\infty}{\lesssim}\frac{R^{slow}_{v\omega}(k,l,y_{1},y_{2})}{s}\mbox{e}^{-\alpha s}\,,\label{eq:estimation-Tvq-bad-case}
\end{equation}
where 
\begin{equation}
R^{slow}_{v\omega}(k,l,y_{1},y_{2})=\frac{\tilde{\omega}_{k,l}^{\infty}(y_{1})\tilde{\omega}_{-k,-l}^{\infty}(y_{2})}{ik(U'(y_{1}))^{2}}
\end{equation}
is a regular function which does not depend on $\alpha$.

If $U(y_{1})\neq U(y_{2})$, the asymptotic oscillations
on the first term in the parenthesis of Eq. (\ref{eq:3.2.3-correlation-vq-explicit-2})
do not cancel out. Using the results of section \ref{sec:appendix-Temporal-decay-of},
we conclude that 
\begin{equation}
\left|\int_{0}^{\infty}\mbox{d}t_1\,\frac{\me^{-i\left[kU(y_{1})-kU(y_{2})\right]t_1-2\alpha t_1}}{(t_1+s)^{2}}\right|\leq\left|\int_{0}^{\infty}\mbox{d}t_1\,\frac{\me^{-i\left[kU(y_{1})-kU(y_{2})\right]t_1}}{(t_1+s)^{2}}\right|\sim\frac{1}{s^{2}}\,.
\end{equation}
For what concerns the second term in the parenthesis of Eq. (\ref{eq:3.2.3-correlation-vq-explicit-2}), we have 
\begin{equation}
\left|\int_{0}^{\infty}\mbox{d}t_1\,\frac{\mbox{e}^{-ikU(y_{1})t_1-2\alpha t_1}}{(t_1+s)^{2}}\tilde{\omega}_{-k,-l}^{r}(y_{2},t_1)\right|\leq\int_{0}^{\infty}\mbox{d}t_1\,\frac{\left|\tilde{\omega}_{-k,-l}^{r}(y_{2},t_1)\right|}{(t_1+s)^{2}}\underset{s\to\infty}{\sim}\frac{A(k,l,y_{2})}{s^{1+\gamma}}\,,
\end{equation}
where $A$ is a positive function which does not depend on $\alpha$.
The formula given above is valid for $0<\gamma<1$ or $\gamma>1$
but not for $\gamma=1$, in which there is a logarithmic correction,
see sections \ref{sec:appendix-Temporal-decay-of}.
The logarithmic correction is not important for the following, so
we do not consider it here.

We thus conclude that, for $U(y_{1})\neq U(y_{2})$
\begin{equation}
\left|T^\alpha_{v\omega}(k,l,y_{1},y_{2},s)\right|\underset{s\to\infty}{\lesssim}\frac{R^{fast}_{v\omega}(k,l,y_{1},y_{2})}{s^{\min\{1+\gamma,2\}}}\mbox{e}^{-\alpha s}
\label{eq:estimation-Tvq-good-case}
\end{equation}
where
\begin{equation}
R^{fast}_{v\omega}(k,l,y_{1},y_{2})=\frac{\tilde{\omega}_{k,l}^{\infty}(y_{1})}{ik(U'(y_{1}))^{2}}A(k,l,y_{2})\,.
\end{equation}

\subsubsection{Large time behaviour of $T^\alpha_{\omega v}$}

We report \eqref{eq:convergence-CR-qv} for convenience,
\begin{equation}
T^\alpha_{\omega v}(k,l,y_{1},y_{2},s)\equiv
\frac12\mathbb{E}_{U}^\alpha\left[\omega_{-k,-l}(y_{1},s)v_{k,l}(y_{2},0)\right]=
\int_{0}^{\infty}\mbox{d}t_1\,\tilde{\omega}_{-k,-l}(y_{1},s+t_1)\tilde{v}_{k,l}(y_{2},t_1)\, .
\end{equation}
We show here that $T^\alpha_{\omega v}$ defined in Eq. (\ref{eq:convergence-CR-qv})
is bounded by a function of $(k,l,y_{1},y_{2})$, independent of~$\alpha$.

We have
\begin{equation}
\left|T^\alpha_{\omega v}(k,l,y_{1},y_{2},s)\right|\leq \mbox{e}^{-\alpha s}|| \tilde{\omega}||_\infty(y_1) \int_{0}^{\infty}\mbox{d}t_1\,\left|\tilde{v}_{k,l}(y_{2},t_1) \right|\,,
\end{equation}
where $||\tilde{\omega}||_\infty = \max_{t_1}\tilde{\omega}_{-k,-l}(y_1,t_1)$ is finite thanks to the Orr mechanism. Using that $\left|\tilde{v}_{k,l}(y_{2},t_1) \right|$ is a bounded function of both $y_2$ and $t_1$, and that it decays as $1/t_1^2$ for $t_1\to\infty$, we conclude that
\begin{equation}
\left|T^\alpha_{\omega v}(k,l,y_{1},y_{2},s)\right|\leq R_{\omega v}(k,l,y_{1},y_{2})\mbox{e}^{-\alpha s}\,.
\label{eq:Tqv-asymptotic-behaviour}\end{equation}
where $R_{\omega v}(k,l,y_{1},y_{2})$ is a positive, bounded function of $(y_1,y_2)$ which does not depend on $\alpha$.

\subsection{Four--points correlation functions\label{sec:SC-background-appendix-4pts}}

\subsubsection{Behavior of $C^\alpha_{kl}$ in the limit $\alpha\to0$}
Using \eqref{eq:Tvv-asymptotic-behaviour} and \eqref{eq:SC-Tqq-computation-appendix} in the definition \eqref{eq:C-vv-appendix},
\[\begin{aligned}
C^\alpha_{kl}(y_{1},y_{2},T_{0})\lesssim \,& \frac{\tilde{\omega}_{-k,-l}^{\infty}(y_{1})\tilde{\omega}_{k,l}^{\infty}(y_{2})R_{vv}(k,l,y_{1},y_{2})}{ik\left[U(y_{1})-U(y_{2})\right]+2\alpha}\int_{T_{0}}^{\infty}\mbox{d}s\,\left(\frac{1}{s^{2}}+o\left(\frac{1}{s^{2}}\right)\right)\me^{-\alpha s}\\
&+R_{vv}(k,l,y_{1},y_{2})\int_{T_{0}}^{\infty}\mbox{d}s\, T_{\omega\omega}^{r,\alpha}(k,l,y_{1},y_{2},s)\, \me^{-\alpha s}\,\left(\frac{1}{s^{2}}+o\left(\frac{1}{s^{2}}\right)\right).\\
\end{aligned}\]
From the properties of $T_{\omega\omega}^{r,\alpha}$, we conclude that
\begin{itemize}
\item if $U(y_1)=U(y_2)$,
\[C^\alpha_{kl}(y_{1},y_{2},T_{0})  \;\underset{\alpha\to0}{=}\;  \frac{A_{1}(k,l,y_1,y_2)}{2\alpha} + o\left(\frac{1}{\alpha}\right)  \;\underset{\alpha\to0}{=}\;  \frac{A_{1}(k,l,y_1,y_2)+o(1)}{2\alpha} \]
where
\[
A_{1}(k,l,y_1,y_2) = \tilde{\omega}_{-k,-l}^{\infty}(y_{1})\tilde{\omega}_{k,l}^{\infty}(y_{2}) \left.\int_{T_0}^\infty \mbox{d}s\, T^\alpha_{vv}(k,l,y_1,y_2,s)\right|_{\alpha=0}\,,
\]
which is finite.
\item if $U(y_1)\neq U(y_2)$,
\[C^\alpha_{kl}(y_{1},y_{2},T_{0})  \;\underset{\alpha\to0}{=}\;  \frac{A_{1}(k,l,y_1,y_2)}{ik\left[U(y_{1})-U(y_{2})\right]} + A_2(k,l,y_1,y_2) \]
where
\[
A_{2}(k,l,y_1,y_2) = \left.\int_{T_0}^\infty \mbox{d}s\, T^\alpha_{vv}(k,l,y_1,y_2,s)T_{\omega\omega}^{r,\alpha}(k,l,y_{1},y_{2},s)\right|_{\alpha=0}\,,
\]
which is finite.
\end{itemize}

\subsubsection{Behavior of $D^\alpha_{kl}$ in the limit $\alpha\to0$}

Using \eqref{eq:estimation-Tvq-bad-case}, \eqref{eq:estimation-Tvq-good-case} and \eqref{eq:Tqv-asymptotic-behaviour} in the definition \eqref{eq:C-vq-appendix}, we have:
\begin{itemize}
\item if $U(y_1)=U(y_2)$,
\[
\left|D^\alpha_{kl}(y_{1},y_{2},T_{0})\right| \lesssim R^{slow}_{v\omega}(k,l,y_{1},y_{2})R_{\omega v}(k,l,y_{1},y_{2})\int_{T_{0}}^{\infty}\mbox{d}s\,\frac{\me^{-2\alpha s}}{s}.
\]
We can now observe that 
\[
\int_{T_{0}}^{\infty}\mbox{d}s\,\frac{\me^{-2\alpha s}}{s}\underset{\alpha\to0}{\sim}\log\alpha T_{0}
\]
so $D^\alpha_{kl}(y_{1},y_{2},T_{0}) = \ln\alpha B_1(k,l,y_{1},y_{2})$ where $B_1$ is finite and doesn't depend on $\alpha$.
\item if $U(y_1)\neq U(y_2)$,
\begin{equation}
\left|D^\alpha_{kl}(y_{1},y_{2},T_{0})\right| \lesssim R^{fast}_{v\omega}(k,l,y_{1},y_{2})R_{\omega v}(k,l,y_{1},y_{2})\int_{T_{0}}^{\infty}\mbox{d}s\,\frac{\me^{-2\alpha s}}{s^{\min{1+\gamma,2}}}.
\end{equation}
We can now observe that 
\begin{equation}
\int_{T_{0}}^{\infty}\mbox{d}s\,\frac{\me^{-2\alpha s}}{s^{\min{1+\gamma,2}}}<\infty\qquad\forall\alpha\geq0
\end{equation}
so  $D^\alpha_{kl}(y_{1},y_{2},T_{0}) = B_2(k,l,y_{1},y_{2})$ where $B_2$ is finite and doesn't depend on $\alpha$.
\end{itemize}

\subsubsection{Conclusion for $\Xi^\alpha_{kl}$}
Collecting the previous results and using $\Xi^\alpha_{kl} = C^\alpha_{kl}+D^\alpha_{kl}$, we have
\begin{itemize}
\item if $U(y_1)=U(y_2)$,
\[\Xi^\alpha_{kl}(y_1,y_2) \underset{\alpha\to0}{=}  \frac{A_{1}(k,l,y_1,y_2)+o(1) + 2\alpha\ln\alpha B_1(k,l,y_{1},y_{2})}{2\alpha}  \underset{\alpha\to0}{=}  \frac{A_{1}(k,l,y_1,y_2)}{2\alpha}. \]
\item  if $U(y_1)\neq U(y_2)$,
\[\Xi^\alpha_{kl}(y_1,y_2) \;\underset{\alpha\to0}{=}\;  \frac{A_{1}(k,l,y_1,y_2) + ik\left[U(y_{1})-U(y_{2})\right]\left[A_2(k,l,y_1,y_2) +B_2(k,l,y_{1},y_{2})\right]}{ik\left[U(y_{1})-U(y_{2})\right]} . \]
\end{itemize}
We conclude that for all $(y_1,y_2)$,
\[\Xi^\alpha_{kl}(y_1,y_2) \underset{\alpha\to0}{\sim}  \frac{A_{kl}(y_1,y_2) }{ik\left[U(y_{1})-U(y_{2})\right]+2\alpha} , \]
with $A_{kl}(y_1,y_2)=A_{1}(k,l,y_1,y_2) + ik\left[U(y_{1})-U(y_{2})\right]\left[A_2(k,l,y_1,y_2) +B_2(k,l,y_{1},y_{2})\right]$.
However, when $U(y_1)=U(y_2)$ and $U'(y_1)=0$, then $A_{kl}(y_1,y_2)=0$. This result indeed follows from the fact that $\tilde{\omega}_{kl}^\infty(y_1)=0$ for such points, see \citep{Bouchet_Morita_2010PhyD} and the discussion in section \ref{sub:Orr}. Then, at such points, $\Xi^\alpha_{kl}(y_1,y_2)$ either converges to a finite value or diverges slower than $1/\alpha$ as $\alpha\to0$. This is the result we wanted to prove and anticipated in eq. (\ref{eq:SC-Xi-kl-convergence}).

\subsection{Temporal decay of some integrals\label{sec:appendix-Temporal-decay-of}}

\subsubsection{Some oscillating integrals}

Consider integrals of the form
\begin{equation}
F(t)=\int_{0}^{\infty}\mbox{d}u\, e^{-igu}f(t+u)\qquad\qquad f(u)\underset{u\to\infty}{\sim}\frac{1}{u^{N}}\qquad\qquad g\neq0\,,
\end{equation}
where $f$ is a smooth real function and $N>0$. We prove here that 
\begin{equation}
F(t)\underset{t\to\infty}{\sim}\frac{1}{t^{N}}.
\end{equation}
Let us perform the change of variable $w=1+u/t$:
\begin{equation}
F(t)=t\, e^{-igt}\int_{1}^{\infty}\mbox{d}w\, e^{-igtw}f(tw)\,=t\, e^{-igt}\int_{1}^{\infty}\mbox{d}w\, e^{-igtw}h_{t}(w)\,,
\end{equation}
where we have introduced the function $h_{t}(w)=f(tw)$; clearly,
$h_{t}(w)\underset{t,w\to\infty}{\sim}\frac{1}{t^{N}w^{N}}$. We also
have $h_{t}^{(n)}(1)\underset{t\to\infty}{\sim}\frac{1}{t^{N}}$ for
all $n$, where $h_{t}^{(n)}$ indicates the $n$-th derivative.

Now perform part integration iteratively on the last expression, for example after two parts integrations:
\begin{equation}
F(t)\underset{t\gg1}{\sim}e^{-igt}\left\{ \frac{-i}{g}h_{t}(1)+\frac{1}{g^{2}t}h_{t}^{(1)}(1)-\frac{1}{g^{2}t}\int_{1}^{\infty}\mbox{d}w\, e^{-igtw}h_{t}^{(2)}(w)\right\} \,.
\end{equation}
Each successive term converges faster to zero than the previous one in the limit $t\ll 1$, thanks to the relation $h_{t}^{(n)}(1)\underset{t\to\infty}{\sim}\frac{1}{t^{N}}$ for
all $n$. We thus have the desired result.

\subsubsection{Non oscillating integrals}

Consider integrals of the form
\begin{equation}
G(t)=\int_{0}^{\infty}\mbox{d}u\,\frac{g(u)}{(u+t)^{2}}\qquad\qquad\int_{0}^{\infty}\mbox{d}u\,g(u)<\infty\,;
\end{equation}
where $g(u)\geq0$ everywhere in $[0,\infty)$. We prove here that
\begin{equation}
G(t)\underset{t\to\infty}{\sim}\frac{A}{t^{2}}\qquad\qquad0<\int_{0}^{\infty}\mbox{d}u\,\frac{g(u)}{(1+u)^{2}}<A<\int_{0}^{\infty}\mbox{d}u\, g(u)\,.\label{eq:behavior-non-oscillating}
\end{equation}
We have
\begin{equation}
G(t)=\frac{1}{t^{2}}\int_{0}^{\infty}\mbox{d}u\,\frac{g(u)}{(1+\frac{u}{t})^{2}};
\end{equation}
let us observe that
\begin{equation}
\frac{1}{(1+u)^{2}}\underset{t>1}{<}\frac{1}{(1+\frac{u}{t})^{2}}<1
\end{equation}
where in the first passage we assumed $t>1$ as we are interested
in the $t\to\infty$ limit of $G$. Then,
\begin{equation}
\frac{1}{t^{2}}\int_{0}^{\infty}\mbox{d}u\,\frac{g(u)}{(1+u)^{2}}\underset{t>1}{<}G(t)<\frac{1}{t^{2}}\int_{0}^{\infty}\mbox{d}u\, g(u)\,.
\end{equation}
We have then proved the desired result in Eq. (\ref{eq:behavior-non-oscillating}). 

These results can be easily extended to the case of integrals of the
form 
\begin{equation}
G(t)=\int_{0}^{\infty}\mbox{d}u\,\frac{g(u)}{(u+t)^{N}}\qquad\qquad\int_{0}^{\infty}\mbox{d}u\,g(u)<\infty\,\qquad\qquad N>0
\end{equation}
and one would obtain the result
\begin{equation}
G(t)\underset{t\to\infty}{\sim}\frac{A}{t^{N}}\qquad\qquad0<\int_{0}^{\infty}\mbox{d}u\,\frac{g(u)}{(1+u)^{N}}<A<\int_{0}^{\infty}\mbox{d}u\, g(u)\,.\label{eq:behavior-non-oscillating-1}
\end{equation}

\subsubsection{Non oscillating integrals where the previous estimation
does not work}

Consider integrals of the form
\begin{equation}
G(t)=\int_{0}^{\infty}\mbox{d}u\,\frac{g(u)}{(u+t)^{2}}\qquad\qquad g(u)\underset{u\to\infty}{\sim}\frac{1}{u^{\gamma}}\qquad\qquad0<\gamma\leq1,
\end{equation}
where $g(u)\geq0$ everywhere in $[0,\infty)$. In this case, the
hypothesis of the previous section do not work because $\int_{0}^{\infty}\mbox{d}u\,g(u)=\infty$.

We prove in this subsection that 
\begin{equation}
G(t)\underset{t\to\infty}{\lesssim}\frac{A_{1}}{t^{1+\gamma}}\qquad\qquad0<\gamma<1\,\label{eq:behavior-non-oscillating-2}
\end{equation}
and
\begin{equation}
G(t)\underset{t\to\infty}{\lesssim}\frac{A_{2}}{t^{2}}\log t\qquad\qquad\textrm{\ensuremath{\gamma}=1}\,\label{eq:behavior-non-oscillating-2-1}
\end{equation}
where $A_{1}$ and $A_{2}$ are suitable positive constants. As usual
the symbol $\underset{t\to\infty}{\lesssim}$ means that there is
a function $G_{1}(t)$ which dominates $G(t)$ and behaves as described
for $t\to\infty$.

The proof of Eq. (\ref{eq:behavior-non-oscillating-2}) and (\ref{eq:behavior-non-oscillating-2-1})
is easily done by observing that $g$ can be majorated for every $u$
by 
\begin{equation}
g(u)\leq\frac{a_{1}}{u^{\gamma}}\qquad\qquad\textrm{if }\qquad0<\gamma<1
\end{equation}
and
\begin{equation}
g(u)\leq\frac{a_{2}}{u+a_{3}}\qquad\qquad\textrm{if }\qquad\gamma=1\,.
\end{equation}
where $a_1$, $a_2$ and $a_3$ are positive constants. The case $0<\gamma<1$ is easily completed by observing that
\begin{equation}
G(t)<G_{1}(t)\equiv\int_{0}^{\infty}\mbox{d}u\,\frac{1}{(u+t)^{2}}\frac{a_{1}}{u^{\gamma}}=a_{1}\pi\left(\frac{1}{t}\right)^{1+\gamma}\gamma\text{Csc}[\pi\gamma]\,\underset{t\to\infty}{\sim}\frac{A_{1}}{t^{1+\gamma}}.
\end{equation}
where Csc is the cosecant%
\footnote{$\text{Csc}(x)<\infty$ if $x\neq n\pi$ with $n$ integer.%
}

The case $\gamma=1$ is also easily accomplished by observing that 
\begin{equation}
G(t)<G_{2}(t)\equiv\int_{0}^{\infty}\mbox{d}u\,\frac{1}{(u+t)^{2}}\frac{a_{2}}{u+a_{3}}=\frac{a_{2}(a_{3}-t-t\ln a_{3}+t\ln t)}{(a_{3}-t)^{2}t}\underset{t\to\infty}{\sim}\frac{A_{2}}{t^{2}}\ln t.
\end{equation}

\bibliographystyle{jfm}
\bibliography{biblio}

\end{document}